\documentclass[11pt,a4paper]{article}

\usepackage{amsmath,amssymb,verbatim}
\usepackage{graphicx,epic,color,float}
\usepackage{overpic} 
\usepackage{enumitem}
\usepackage[round]{natbib}
\usepackage[english]{babel}
\usepackage[T1]{fontenc}
\usepackage[utf8]{inputenc}
\usepackage[labelfont=bf,labelformat=simple]{caption,subcaption}
\usepackage{lmodern}
\usepackage{hyperref}
\usepackage{geometry}
\usepackage{ragged2e}
\usepackage{setspace}
\usepackage{upgreek,bbm}
\usepackage{siunitx}
\usepackage{gensymb}
\usepackage{tabularx}
\usepackage{framed} 
\usepackage{longtable}
\usepackage{contour} 
\usepackage{multirow} 
\usepackage[affil-it]{authblk}
\usepackage{pgfplots}
\pgfplotsset{compat=newest}
\pgfplotsset{scaled y ticks=false}
\graphicspath{{.}{./images/}}
\usepackage{tikz}
\usetikzlibrary{plotmarks,spy,calc,external,through,backgrounds,shapes.geometric}

\graphicspath{{.}{./figures/}}
\definecolor{rwthblue}{rgb}{0,0.33,0.62}
\definecolor{rwthmedblue}{rgb}{0.25,0.5,0.72}
\definecolor{rwthlightblue}{rgb}{0.56,0.73,0.9}

\definecolor{rwthturq}{rgb}{0,0.6,0.63}
\definecolor{rwthlightturq}{rgb}{0.54,0.8,0.81}
\definecolor{rwthlila}{rgb}{0.48,0.44,0.67}
\definecolor{rwthgreen}{rgb}{0.34,0.67,0.15}
\definecolor{rwthlightgreen}{rgb}{0.72,0.84,0.6}
\definecolor{rwthred}{rgb}{0.8,0.03,0.12}
\definecolor{rwthyellow}{rgb}{1,0.93,0}
\definecolor{rwthpurple}{rgb}{0.38,0.13,0.35}
\definecolor{rwthlightpurple}{rgb}{0.66,0.52,0.62}
\definecolor{darkblue}{rgb}{0,0,1}
\definecolor{rwthorange}{rgb}{0.96,0.66,0}
\hypersetup{colorlinks=true, breaklinks=true, linkcolor=darkblue, menucolor=darkblue, citecolor=darkblue, urlcolor=darkblue}

\newcommand{\mr}[1]{\text{#1}}

\newcommand{\mmax}{\mathrm{max}}

\begin{document}
%
\title{\Large{\bf{Modeling the debonding process of osseointegrated implants due to coupled adhesion and friction}}}

\author[1,2]{Katharina Immel} 
\author[2]{Vu-Hieu Nguyen} 
\author[2]{Guillaume Ha\"{i}at}
\author[1,3,4\footnote{Corresponding author, email: sauer@aices.rwth-aachen.de}]{\\Roger A. Sauer}

\affil[1]{Aachen Institute for Advanced Study in Computational Engineering Science (AICES), RWTH Aachen University, Templergraben 55, 52056 Aachen, Germany}
\affil[2]{MSME, CNRS UMR 8208, Université Paris-Est Créteil, Université Gustave Eiffel, \linebreak 94010 Créteil, France}
\affil[3]{Faculty of Civil and Environmental Engineering, Gdańsk University of Technology, \linebreak ul. Narutowicza 11/12, 80-233 Gdańsk, Poland}
\affil[4]{Department of Mechanical Engineering, Indian Institute of Technology Guwahati, Assam 781039, India}
\vspace{-10mm}
\date{\small{Published\footnote{This PDF is the personal version of an article whose final publication is available at \href{https://link.springer.com/article/10.1007/s10237-022-01637-7}{https://link.springer.com/}} $\,$ in \textit{Biomechanics and Modeling in Mechanobiology}
\href{https://link.springer.com/article/10.1007/s10237-022-01637-7}{DOI: 10.1007/s10237-022-01637-7} \\
Submitted on  20 March 2022, Revised on 26 August 2022, Accepted on 6 September 2022}}
\maketitle
\vspace{-12mm}
%
\rule{\linewidth}{.15mm}
{\bf Abstract:}
Cementless implants have become widely used for total hip replacement surgery.
The long-term stability of these implants is achieved by bone growing around and into the rough surface of the implant, a process called osseointegration.
However, debonding of the bone--implant interface can still occur due to aseptic implant loosening and insufficient osseointegration, which may have dramatic consequences.
The aim of this work is to describe a new 3D finite element frictional contact formulation for the debonding of partially osseointegrated implants.
The contact model is based on a modified Coulomb friction law~\citep{immel2020} that takes into account the tangential debonding of the bone-implant interface. 
This model is extended in the direction normal to the bone-implant interface by considering a cohesive zone model, to account for adhesion phenomena in the normal direction and for adhesive friction of partially bonded interfaces.
The model is applied to simulate the debonding of an acetabular cup implant.
The influence of partial osseointegration and adhesive effects on the long-term stability of the implant is assessed. 
The influence of different patient- and implant-specific parameters such as the friction coefficient $\mu_\text{b}$, the trabecular Young's modulus $E_\text{b}$ and the interference fit $I\!F$ is also analyzed, in order to determine the optimal stability for different configurations.
Furthermore, this work provides guidelines for future experimental and computational studies, that are necessary for further parameter calibration.

{\bf Keywords:}	adhesion, bone--implant interface, debonding, friction, nonlinear finite element analysis, osseointegration \\
\rule{\linewidth}{.15mm}
%
\section{Introduction} \label{s:intro} 	
Uncemented acetabular cup implants have become the gold standard for acetabular replacement~\citep{small2013} in the context of total hip arthroplasty~\citep{philippot2009}.
However, aseptic implant loosening is a major complication of total hip arthroplasty and is mostly due to issues related to implant stability.
The primary (or initial) stability of acetabular cup implants is established during the surgery and is governed by mechanical factors, such as surface interlocking and bone quality.
Secondary (or long-term) stability is achieved several weeks or months after surgery, through the formation and maturation of newly formed bone tissue at the bone-implant interface, a process called \textit{osseointegration}~\citep{albrektsson1981}. 
While the evolution of secondary implant stability is governed by complex biochemical processes, the mechanical behavior of the bone-implant interface remains crucial for the surgical outcome~\citep{gao2019}.

Pull--out, lever--out, and torsional debonding tests have been widely used to evaluate the implant fixation by recording the force--displacement curve, the maximum removal force or the shear strength of the bone-implant interface~\citep{soballe1993,brunski2000,chang2010,trisi2011,mathieu2012}, which have been correlated with histological assessments in animal studies~\citep{johansson1991,haiat2014}.
A simple animal model consisting in considering coin-shaped implants has been used by different groups~\citep{ronold2002,mathieu2012,gao2019} in order to investigate the evolution of the secondary stability of osseointegrated implants.
The interest of such an approach is to work in a standardized configuration with a planar bone-implant interface, allowing to carry out efficient and precise experimental testing. 
However, the influence of biological as well as mechanical factors on the long--term stability and the size and shape of joint implants, makes experimental testing of this kind of implants difficult and at present, such studies are lacking in the literature~\citep{viceconti2004,helgason2008}.
Therefore, there is a high demand in reliable numerical models that can predict long--term stability of implants to aid in implant conception but also to chose the best implantation technique in a patient specific manner.
The advantage of numerical models is that all parameters can be precisely controlled, which is not the case when working with animal models.
The difficulty of predicting implant stability arises from the complex nature of the bone-implant interface, which is related to (1) the implant surface roughness, (2) the non-homogeneous contact between bone and implant, (3) local adhesion and friction, (4) the time dependence of peri-implant bone properties~\citep{mathieu2012}, and (5) the widely varying loading scenarios during the implant life cycle.
While the implant material, roughness, and surface coating are important factors determining implant stability, friction phenomena also play a major role and, in turn, depend on the surface roughness and bone quality~\citep{shirazi-adl1993}. 

Most finite element (FE) models that study implant behavior assume the bone-implant interface to be either perfectly bonded or fully sliding~\citep{gupta2010,tomaszewski2010,chou2014,demenko2016,rittel2017,mondal2019}. While perfectly bonded contact conditions do not allow for debonding to occur, interface behavior that is modeled as frictionless or by classical Coulomb's friction cannot fully represent the non--linear interface behavior of the bone-implant interface before nor after osseointegration occurs~\citep{dammak1997b,viceconti2004}.
Furthermore, it was shown that the implants are never fully osseointegrated, rather the bone--to--implant contact ratio is typically only between 30 and 70~\% after healing~\citep{branemark1997,marin2010}.
Therefore, imperfect osseointegration and its influence on the evolution of implant stability must be considered.
A common approach is to model imperfect osseointegration by setting osseointegrated contact regions to be perfectly bonded, while not integrated contact regions follow frictionless or Coulomb's friction contact~\citep{spears2001,viceconti2004,helgason2009,galloway2013}. 
Another approach is to consider a varying degree of osseointegration and to adjust the material properties of the bone-implant interface, while keeping the interface fully bonded~\citep{kurniawan2012} or by varying the friction coefficient of the bone-implant interface from $\mu=0$ for fully unbonded to $\mu=\infty$ for perfectly osseointegrated surfaces~\citep{korabi2017}.
However, although interesting, these models cannot represent the adaptive changes of the bone-implant interface, and debonding is usually characterized by excessive stress or strain at the bone-implant interface or in the bone, without modeling the actual separation between bone and implant and local changes of contact conditions.

To the best of the authors’ knowledge, there is currently no other continuum mechanics model that i) takes into account friction and adhesion in normal and tangential direction without using fully bonded interfaces, ii) takes into account imperfect/partial osseointegration, iii) models adhesive debonding in normal and tangential direction, and iv) was used to model implant debonding on the macroscopic scale.
So far, there is only one similar model we know of, which is the cohesive debonding model of~\cite{rittel2018} that was used to model the debonding of partially integrated dental implants.
There, a tie constraint was applied to the bone-implant interface, so that cohesive failure occurs in the bone tissue around the bone-implant interface.
Partial osseointegration was modeled by defining a relative osseointegrated area with random distribution and considering non-integrated areas in frictional contact. 
However, comparison is difficult as there are considerable differences between adhesive and cohesive debonding, the choice of geometries and contact conditions.

The aim of this work is to propose a phenomenological model for the frictional contact behavior of debonding osseointegrated implants.
In previous work by our group~\citep{immel2020}, the adhesive failure and tangential debonding of partially osseointegrated coin-shaped implants was modeled by proposing a modified Coulomb's law calibrated from experimental data.
In this work, this modified Coulomb's friction law for tangential debonding, is first extended by a cohesive zone model in order to account for debonding in the normal direction as well as adhesive friction.
Second, this updated contact model is demonstrated on an osseointegrated, coin-shaped implant using mode I, III, and mixed mode debonding. Third, the modified Coulomb's friction law is then applied to simulate the debonding of a 3D, osseointegrated acetabular cup implant in different removal tests.
The implant stability is quantified by assessing the removal force/torque and the biomechanical determinants of the long-term stability, such as primary stability and degree of osseointegration, are assessed.

The remainder of this work is structured as follows: 
Section~\ref{s:model} describes the governing equations and the contact formulation for normal and tangential debonding.
In Section~\ref{s:csi_validation}, the demonstration with a coin-shaped implant is presented. 
The extended contact model is then applied to simulate removal experiments of osseointegrated acetabular cup implants in Section~\ref{s:ACI2_ACI2}.
Section~\ref{s:results_disc} discusses the results obtained with the modified Coulomb's law and its new extension and provides perspectives and guidelines for future experimental and numerical studies.
Last, Section~\ref{s:conclusion} gives a conclusion and an outlook to possible extensions.

\section{Models and Methods} \label{s:model}
This section discusses the governing equations describing the contact behavior, including a summary of the modified Coulomb's friction law~\citep{immel2020} and its extension in normal direction.
The resulting equations are discretized within a finite element framework to obtain a numerical solution.
The readers are referred to \cite{sauer2013,sauer2015}, \cite{duong2019} and \cite{immel2020} for a more detailed derivation of the considered contact formulation and its finite element implementation.
%
\subsection{Material law} \label{s:model_material}
Throughout this work, a hyper-elastic Neo-Hookean material model is used for all bodies.
The stress--strain relation for the Cauchy stress $\boldsymbol{\sigma}$ of this model is defined by~\citep{zienkiewicz2005}
\begin{equation} \label{eq:neohooke}
	\boldsymbol{\sigma} = \frac{\Lambda}{J} \left( \mathrm{ln} J \right) \boldsymbol{I} + \frac{G}{J} \left( \boldsymbol{b} - \boldsymbol{I} \right), 
\end{equation}
with the volume change $J$, the identity tensor $\boldsymbol{I}$, and the left Cauchy-Green tensor $\boldsymbol{b}$.
The Lamé parameters $G$ (shear modulus) and $\Lambda$ can be expressed in terms of Young's modulus $E$ and Poisson ratio $\nu$, by
\begin{equation}\label{eq:young}
	G = \frac{E}{2(1+\nu)} \quad \mathrm{and} \quad \Lambda = \frac{2G \nu}{1 - 2\nu}.
\end{equation}
%
\subsection{Contact kinematics and contact traction} \label{s:model_kinematics}		
The contact traction $\boldsymbol{t}_\text{c}$ can be decomposed into a normal and tangential component, i.e.
\begin{equation}
	\boldsymbol{t}_\text{c} = \boldsymbol{t}_\text{n} + \boldsymbol{t}_\text{t}.
\end{equation}
The normal traction is proportional to the negative contact pressure $p$, i.e.
\begin{equation} \label{eq:normal_traction}
	\boldsymbol{t}_\text{n} = -p \, \boldsymbol{n}
\end{equation}
where $\boldsymbol{n}$ is the outward surface normal.
The contact pressure $p$ can be modeled by the penalty method, i.e.
\begin{equation}
	p = -\epsilon_\text{n} g_\text{n}, \quad  
\end{equation}
which is active when the bodies penetrate, i.e. the normal gap $g_\text{n} = \boldsymbol{g}_\text{n}\cdot\boldsymbol{n}$  becomes negative ($g_\text{n} < 0$).
Here, $\boldsymbol{g}_\text{n}$ is the normal contact gap vector between surface points on the two bodies, and $\epsilon_\text{n}$ is the penalty parameter.

For frictional contact, the tangential traction is determined by the behavior during sticking and sliding, and the distinction between these two states is based on a slip criterion of the form
\begin{equation}
	f_\text{s} \begin{cases}
		<0, & \text{for sticking},\\
		=0, & \text{for sliding}.
	\end{cases}
\end{equation}
It can be formulated as
\begin{equation} \label{eq:slip_criterion}
	f_\text{s} = \lVert \boldsymbol{t}_\text{t} \rVert - t^\text{slide}_\text{t},
\end{equation}
where $t^\text{slide}_\text{t}>0$ is a limit value for the tangential traction that corresponds to the magnitude of the tangential traction during sliding.
For sticking, the tangential traction is defined by the constraint that no relative tangential motion occurs, while for sliding, the tangential traction is defined by a sliding law.
One of the simplest but most commonly used sliding laws is Coulomb's friction law, which states
\begin{equation}
	t^\text{slide}_\text{t} =  \mu \, p, \label{eq:coulomb}
\end{equation}
where the friction coefficient $\mu$ is a material parameter that can change locally ($\mu = \mu(\boldsymbol{x})$), but does not depend on $p$ or $g_\text{n}$.
%
\subsubsection{State function} \label{s:model_state}
To model the current bonding state at each point $\boldsymbol{x}$ on the interface and its transition from fully bonded to fully broken, we use a smooth state function $\phi$~\citep{immel2020}. 
This state function depends on the initial bonding state $\phi_0$, the model parameters $a_\text{s}, b_\text{s}$ and a damage parameter $g_\text{d}(\boldsymbol{x})$, i.e.
\begin{equation} \label{eq:phi}
	\displaystyle 
	\begin{aligned} 
		&\phi(\boldsymbol{x}, g_\text{d}(\boldsymbol{x})) = \phi_0(\boldsymbol{x}) \cdot  \\  &\begin{cases}
			1, & c_1 < 1, \\
			\frac{1}{2} - \frac{1}{2} \mathrm{sin} \left( \frac{\pi}{2b_\text{s}} \left( c_1 - b_\text{s} - 1 \right) \right), 
			& 1 \le c_1 \le c_2,\\
			0, & c_1 > c_2,
		\end{cases}
	\end{aligned} 
\end{equation}
where $c_1=\frac{g_\text{d}(\boldsymbol{x})}{a_\text{s}}$ and $c_2=1+2b_\text{s}$.
Here, $a_\text{s}$ denotes a displacement threshold, while $b_s$ defines the size of the transition zone from a fully bonded to a fully debonded state.
The initial bonding state $\phi_0$ can range between 0 (no initial bonding) and 1 (full initial bonding).
A value in between indicates imperfect osseointegration, i.e. $\phi_0=0.5$ denotes that the level of osseointegration has reached 50\%.

\begin{figure}[H]
	\centering
	\includegraphics[width=0.5\linewidth]{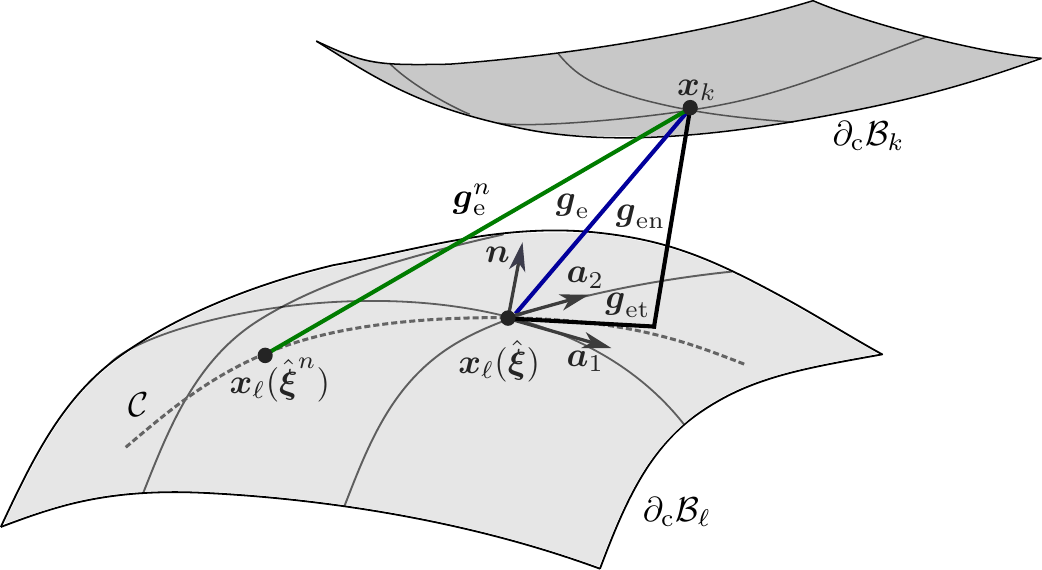}
	\caption{Frictional contact kinematics at current time $t_{n+1}$: current slave point $\boldsymbol{x}_k$, current position of the previous interacting point $\displaystyle \boldsymbol{x}_\ell(\hat{\boldsymbol{\xi}}^n)$, current interacting point $\boldsymbol{x}_\ell(\hat{\boldsymbol{\xi}})$, the current elastic gap vector $\boldsymbol{g}_\text{e}$ and its normal and tangential components, previous elastic gap $\boldsymbol{g}^n_\text{e}$, and the sliding path $\mathcal{C}$. Adopted from~\cite{immel2020}.}
	\label{img:unified_formulation}
\end{figure}

The damage parameter $g_\text{d}(\boldsymbol{x})$ is chosen to be additively composed of an accumulated irreversible tangential slip $g_\text{s}$ and the accumulated normal gap $g_\text{sn}$, i.e.
\begin{equation} \label{eq:accumulated_def}
	g_\text{d}= g_\text{s} + g_\text{sn}.
\end{equation}
Numerically, $g_\text{s}$ and $g_\text{sn}$ can be approximated as (see Fig.~\ref{img:unified_formulation} and~\cite{immel2020}, Appendix A.2)
\begin{equation}
	\begin{aligned}
		g_\text{s}^{n+1} \approx& \sum_{i=1}^{n+1} \lVert \boldsymbol{x}_\ell \left( \hat{\boldsymbol{\xi}}^i \right) - \boldsymbol{x}_\ell \left( \hat{\boldsymbol{\xi}}^{i-1} \right) \rVert,\\
		g_\text{sn}^{n+1} \approx& \sum_{i=1}^{n+1} \lVert g^{i}_\text{en} - g^{i-1}_\text{en} \rVert, 
	\end{aligned}
\end{equation}
where
\begin{equation}
	\begin{aligned}
		g^i_\text{en} &:= \boldsymbol{g}^i_\text{en} \cdot \boldsymbol{n}, \\
		\boldsymbol{g}^i_\text{en} &:= (\boldsymbol{n} \otimes \boldsymbol{n}) \, \boldsymbol{g}^i_\text{e},\\ \boldsymbol{g}^i_\text{e} &:= \boldsymbol{x}^i_k - \boldsymbol{x}^i_\ell(\hat{\boldsymbol{\xi}}).
	\end{aligned}		
\end{equation}
Here, subscripts $i$ and $i-1$ denote the current and previous quantities at time step $t_i$, analogously to the quantities at $t_n+1$ shown in Figure 1.
By definition, $g_\text{s}$ and $g_\text{sn}$ are always positive and monotonically increasing, even when the loading reverses.
The modified Coulomb's law only depends on $g_\text{s}$, while the extension also takes into account $g_\text{sn}$.
This implies that during sticking, $g_\text{d}$ only increases if $g_\text{sn}$ increases. Otherwise, there is no change in the debonding state.
As this is a local model, where $\phi$ can change pointwise, it allows for the description of locally varying bonding states, such as occur in crack propagation and partial osseointegration.
According to Eq.~\eqref{eq:phi}, the state variable $\phi$ determines whether a point is in an unbroken ($\phi=1$), partially broken ($0<\phi<1$) or fully broken state ($\phi=0$).
The current bonding state determines the contact traction components, which is explained in the following.
%
\subsubsection{Adhesive friction and debonding} \label{s:model_adhesion}
\begin{figure}[H]
	\centering
	\begin{subfigure}[t]{0.41\linewidth}
		\centering
\includegraphics[width=\linewidth]{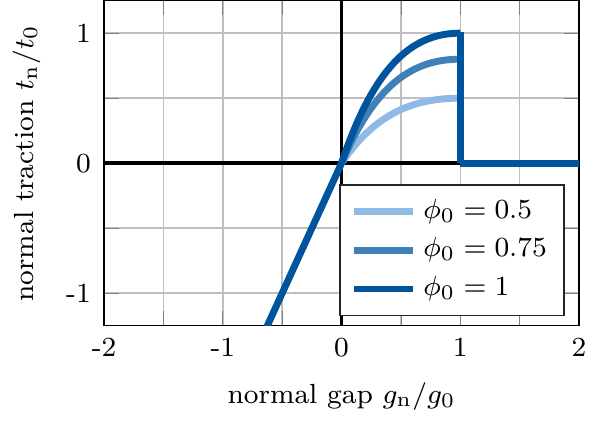}
		\caption{Exponential cohesive zone model based on the normal gap $g_\text{n}$ for various $\phi_0$, $g_\text{b}=g_0$, and $t_\text{n} := \boldsymbol{t}_\text{n}\cdot\boldsymbol{n}.$}
		\label{img:normal_traction}
	\end{subfigure}
	\hspace{5mm}
	\begin{subfigure}[t]{0.4\linewidth}
		\centering
	\includegraphics[width=\linewidth]{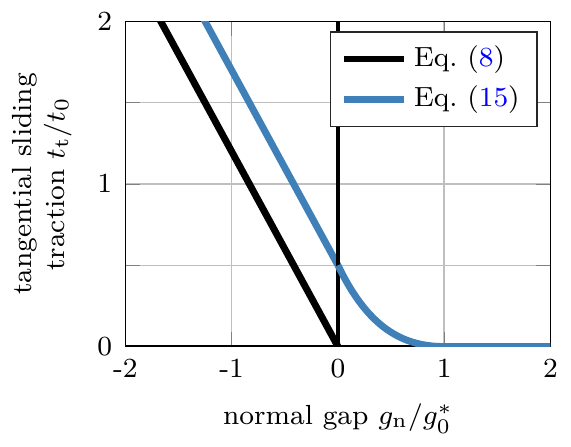}
		\caption{Tangential sliding traction $t_\text{t}$ as a function of the normal gap $g_\text{n}$.}
		\label{img:friction cone}
	\end{subfigure}
	\caption{Illustration of cohesive zone model (a) and the extended modified Coulomb's law with adhesive friction (b) for $\phi_0 > 0$.}
	\label{img:model}
\end{figure}

To account for normal adhesion and debonding in the extension of the modified Coulomb's law, the normal traction~\eqref{eq:normal_traction} is extended by an exponential cohesive zone model (CZM)~\citep{xu1992} (see \cite{sauer2016}), i.e.
\begin{equation} \label{eq:adhesion_traction}
	\displaystyle
	\begin{aligned}
		&	\boldsymbol{t}_\text{n} = \\ &\begin{cases}
			\boldsymbol{0},  &    g_\text{b} \ge g_\text{n} \text{ or } \phi=0,\\
			\frac{\phi_0 t_0 g_\text{n}}{g_0} \,\text{exp}\left(1-\frac{g_\text{n}}{g_0}\right) \boldsymbol{n}, & 0 \le g_\text{n}<g_\text{b} \text{ and } \phi > 0, \\
			-\epsilon_\text{n} \, \boldsymbol{g}_\text{n}, & g_\text{n} < 0,
		\end{cases}
	\end{aligned}
\end{equation}
where $t_0$ is the maximum positive normal traction, $g_0$ is the contact distance, where the maximum traction $t_0$ occurs, and $g_\text{b}$ is a cut-off distance, where contact is lost.
The parameters $t_0, g_0, g_\text{b}$ depend on the interface.
The normal traction model~\eqref{eq:adhesion_traction} is illustrated in Figure~\ref{img:normal_traction}.

Equation~\eqref{eq:adhesion_traction} implies that, when pulling the contact surfaces apart in normal direction, as long as the point remains fully or partially bonded ($\phi > 0$) the normal traction keeps increasing until $g_\text{n} = g_\text{b}$. 
As soon as the point is fully debonded ($\phi=0$) or the normal gap is $g_\text{n}>g_\text{b}$, the contact is lost and the normal traction component becomes $\boldsymbol{t}_\text{n}=\boldsymbol{0}$.
The sharp drop in the normal traction at $g_\text{b}$ is motivated by observations from experimental pull-out tests of osseointegrated, coin-shaped implants~\citep{ronold2002,nonhoff2015}.

\begin{figure}[H]
	\centering
	\includegraphics[width=0.4\linewidth]{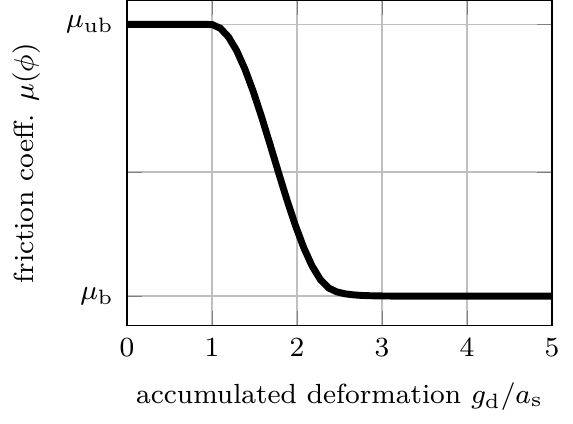}
	\caption{Illustration of the friction function $\mu(\phi)$ based on Eqs.~\eqref{eq:mod_mu} and \eqref{eq:phi} with respect to the accumulated deformation $g_\text{d}$. Parameter $a_\text{s}$ denotes a displacement threshold.}
	\label{img:mu}
\end{figure}

In this work, the modified Coulomb's law introduced by~\cite{immel2020} is used to model the tangential debonding of osseointegrated interfaces.
There, the friction coefficient $\mu$ is modeled as a function of the scalar state variable $\phi$, as
\begin{equation} \label{eq:mod_mu}
	\mu := \mu(\phi) = \phi \, \mu_\text{ub} + \left( 1 - \phi \right) \mu_\text{b}, 
\end{equation}
where $\mu_\text{ub}$ and $\mu_\text{b}$ are the friction coefficient for the unbroken (initial) and broken state, respectively, that are weighted according to the state variable.
The friction function~\eqref{eq:mod_mu} is illustrated in Figure~\ref{img:mu}.
To enable sliding for tensile normal traction in the present extension of the contact model, the tangential sliding limit is shifted by
\begin{equation} \label{eq:adhesive_friction}
	t^\text{slide}_\text{t} = \mu(\phi) \left( t_0 - t_\text{n}\right),
\end{equation}
according to~\cite{mergel2019,mergel2021}.

The slope of the function $t_\text{n}(g_\text{n})$ and $t^\text{slide}_\text{t}(\mu)$ at $g_\text{n}=0$ depends solely on the choice of the parameters.
It is smooth when
\begin{equation} \label{eq:smooth}
	\epsilon_\text{n} = \phi_0 \, e \frac{t_0}{g_0},
\end{equation}
(where $e$ is Euler's number), otherwise it is discontinuous.
A comparison of the standard Coulomb's law and the proposed extended modified Coulomb's law based on adhesive friction is shown in Figure~\ref{img:model}.
A list of all constant parameters of the modified Coulomb friction law with their respective mode of determination is given in Table~\ref{tab:model_params}.

\begin{table}[H]
	\centering
	\begin{tabular}{ccll}
		\hline 
		symbol & unit & parameter & method of determination\\ 
		\hline 
		$a_\text{s}$ & mm & displacement threshold & parameter study\\
		\multirow{2}{*}{$b_\text{s}$}& \multirow{2}{*}{--} & size of transition zone state & \multirow{2}{*}{parameter study}\\
		& & between fully bonded and fully broken state & \\
		$\phi_0$ & -- &  initial degree of osseointegration & experimentally\\
		$\mu_\text{ub}$ & -- & friction coefficient for unbroken state &  parameter study\\
		$\mu_\text{b}$ & -- & friction coefficient for broken state & experimentally\\
		$t_0$ & [Pa] & maximum positive traction & experimentally \\
		$ g_0$ & mm & contact distance where $t_0$ occurs & experimentally\\
		$g_\text{b}$ & mm & cut-off distance where normal contact is lost & experimentally\\
		\hline
	\end{tabular}
	\caption[]{List of the constant parameters of the modified Coulomb friction law with their mode of determination.}
	\label{tab:model_params}
\end{table}

%
\section{Application to coin-shaped implants} \label{s:csi_validation}
To demonstrate the new contact formulation~\eqref{eq:adhesion_traction} and~\eqref{eq:adhesive_friction} a simple implant model of an osseointegrated coin-shaped implant (CSI)~\citep{mathieu2011,vayron2012,mathieu2012,mathieu2012b,vayron2014,fraulob2020a} is used to simulate different debonding modes.
%
\subsection{Setup} \label{s:csi_setup}
We consider the same basic setup as in~\cite{immel2020}, which is briefly summarized below.
The CSI is a short cylinder with radius $R_\text{i}=2.5$ mm and height $H_\text{i}=3$ mm.
The bone sample is modeled as a rectangular cuboid with dimensions 12.5 $\times$ 12.5 $\times$ 5 mm. 
The implant is positioned at the center of the upper bone surface.
For both bodies, the Neo-Hookean material model of Eq.~\eqref{eq:neohooke} is used.
The material properties for the implant are those of titanium alloy (Ti-6Al-4V; $E_\text{i}=113$ GPa, $\nu_\text{i}=0.3$).
The elastic properties of the bone block are $E_\text{b}=18$ GPa, $\nu_\text{b}=0.3$.
Furthermore, all materials are assumed to be homogeneous and isotropic and both contact surfaces are assumed to be perfectly flat.

The bodies are meshed according to the parameters given in Table \ref{tab:csi_mesh}, where $n_\text{el}$ denotes the number of elements of the body/surface and $n_\text{gp}$ denotes the number of Gauss points per element.
While the bulk is discretized with linear Lagrange shape functions, the contact surfaces are discretized with enriched contact finite elements based on quadratic non-uniform, rational B-splines (NURBS)~\citep{corbett2014,corbett2015}.

The parameters of the state function~\eqref{eq:phi} are chosen to be $a_\text{s}=22$ µm, $b_\text{s}=0.74$, $\mu_\text{ub}=0.44$, and $\mu_\text{b}=0.3$, based on~\cite{immel2020}.
The initial osseointegration is constant across the bone-implant interface and is set to be $\phi_0=1$ (perfectly integrated).
Due to the lack of experimental data, the cohesive zone parameters $g_0, g_\text{b}$ are set to $g_\text{b}=g_0=3a_\text{s}$, for simplicity.
\begin{table}[H]
	\centering
	\begin{tabular}{lccc}
		\hline 
		body & $n_\text{el}$ & shape fcts. & $n_\text{gp}$\\ 
		\hline 
		implant bulk & 18 & linear Lagrange & $2\times 2\times 2$ \\
		bone bulk & 450 & linear Lagrange & $2\times 2\times 2$ \\
		lower implant surface & 9 & quad. NURBS & $5\times 5$ \\
		upper bone surface & 225 & quad. NURBS & $5\times 5$\\
		\hline
	\end{tabular} 
	\caption[]{CSI debonding: Number of finite elements $n_\text{el}$, type of shape functions and number of Gauss points per element $n_\text{gp}$ for the two bodies and their contact surfaces. Adopted from~\cite{immel2020}.}
	\label{tab:csi_mesh}
\end{table}
\begin{table}[H]
	\centering
	\begin{tabular}{ccc}
		\hline 
		parameter & value & source \\ 
		\hline 
		$a_\text{s}$ & 22 µm & \cite{immel2020} \\
		$b_\text{s}$ & 0.74 & \cite{immel2020} \\
		$\phi_0$ & 1 & assuming perfect bonding \\
		$\mu_\text{ub}$ & 0.44 &  \cite{immel2020} \\
		$\mu_\text{b}$ & 0.3 &  \cite{immel2020} \\
		$t^*_0$ & 1.8 MPa & \cite{ronold2002} \\
		$g_0$ & 3$a_\text{s}$ & chosen reasonably\\
		$g_\text{b}$ & $a_\text{s}$ & chosen reasonably\\
		\hline
	\end{tabular}
	\caption[]{List of the constant parameters of the modified Coulomb friction law and the chosen values.}
	\label{tab:csi_params}\end{table}
The maximum traction of the cohesive zone model, $t^*_0=1.8$ MPa, is calibrated based on the results of~\cite{ronold2002} for polished, titanium CSIs. 
In that experimental study, CSIs with different surface roughness were implanted into rabbit tibiae and allowed to osseointegrate for 10 weeks. 
Then, the implants were removed together the surrounding bone.
The bone and implant parts were fixed into a tensile test machine and the implant was pulled constantly in the normal direction until it was completely debonded from the bone.
For the polished CSI an average degree of osseointegration of $\phi_0=0.26$ and an average maximum pull-out force of 9~N were determined, which results in an approximate 35~N for $\phi=1$. 
All parameters of the contact model and their values are listed in Table~\ref{tab:csi_params}.

\begin{figure}[H]
	\centering
	\includegraphics[width=0.75\linewidth]{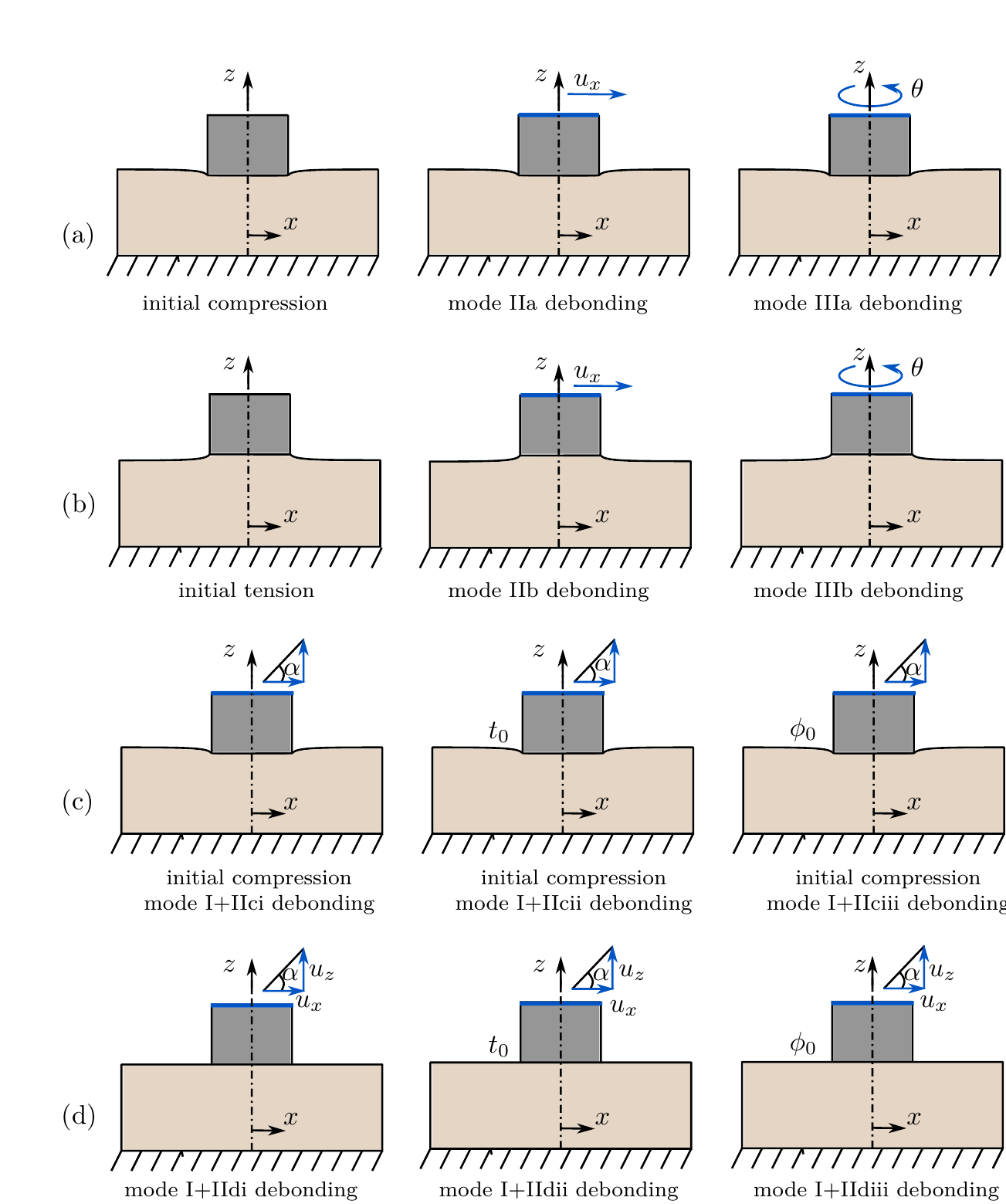}
	\caption[]{CSI debonding: Illustrations of the boundary conditions for different debonding cases. (a) Debonding under an initial compression of -70 N, either in mode II or III. (b) Debonding under an initial tension of 20 N, either in mode II or III. Mixed mode debonding with initial compression of -70 N (ci) and without initial contact force (di), and under various loading angles $\alpha$. Mixed mode debonding with initial compression of -70 N (cii) and without initial contact force (dii), under loading angle $\alpha=45^\circ$ and various CZM parameter $t_0$ in Eq.~\eqref{eq:adhesion_traction}. Mixed mode debonding with initial compression of -70 N (ciii) and without initial contact force (diii), under loading angle $\alpha=45^\circ$ and various initial degrees of osseointegration $\phi_0$. Altogether five test cases are investigated as discussed in the text.}
	\label{img:aci2_csi_bc}
\end{figure}

The boundary conditions and considered test configurations are shown in Figure~\ref{img:aci2_csi_bc}.
The lower surface of the bone block is fixed in all directions.
In this work, only quasi-static conditions are considered.
\pagebreak
First, the implant is pressed into the bone block until a normal reaction force of -70 N is reached, as is done in~\cite{immel2020} for the corresponding parameter set $a_\text{s}=22$ µm, $b_\text{s}=0.74$, $\mu_\text{ub}=0.44$, and $\mu_\text{b}=0.3$. 
Then, for the first three test cases, full and homogeneous initial osseointegration ($\phi_0=1$) of the bone-implant interface is applied (Fig.~\ref{img:aci2_csi_bc}(a) and (c)).
For test cases with tension, the implant is then pulled in normal direction until an average normal reaction force of 20 N is reached (Fig.~\ref{img:aci2_csi_bc}(b)).
Last, debonding with no initial pressure or tension is considered (Fig.~\ref{img:aci2_csi_bc}(d)).
Then, the new contact model is examined for five different debonding test cases:
\begin{enumerate}
	\item mode II: the upper implant surface is moved in $x$-direction under constant compression (mode IIa) or tension (mode IIb).
	\item mode III: the upper surface of the implant is rotated around its $z$-axis under constant compression (mode IIIa) or tension (mode IIIb).
	\item mode I+II: the upper implant surface simultaneously pulled along the $z$-axis and in $x$-direction, corresponding to an angle $\alpha=30, 45,$ or 60$^\circ$. This is performed with initial compression (mode I+IIci) and without initial contact force (mode IIIdi).
	\item mode I+II: the upper implant surface is simultaneously pulled along the $z$-axis and in $x$-direction, corresponding to an angle of $\alpha=45^\circ$ for different choices of $t_0\in[t^*_0/2, t^*_0, 2t^*_0]$. This is performed with initial compression (mode I+IIcii) and without initial contact force (mode I+IIdii).
	\item mode I+II: the upper implant surface is simultaneously pulled along the $z$-axis and in $x$-direction, corresponding to an angle of $\alpha=45^\circ$ for increasing degrees of initial osseointegration $\phi_0\in[0, 0.25, 0.5, 0.75, 1]$. This is performed with initial compression (mode I+IIciii) and without initial contact force (mode I+IIdiii).
\end{enumerate}
All simulations are performed with an in-house, MATLAB-based solver (R2019b, The MathWorks, Natick, MA, USA). 
Contact is computed with a penalty regularization, and the corresponding penalty parameter is set to $\epsilon_\text{n}=\epsilon_\text{t}= E_\text{b}/L_0$, with $L_0=0.01$ m.
The step size for all load cases is $\Delta u=0.65$ µm (for applied displacement loads and $\Delta \theta=0.1^\circ$ for applied rotations).


%
\subsection{Results}
In the following, the results of the debonding tests for the CSI, in terms of load-displacement curves, obtained with the modified Coulomb's law (MC) and its new extension to adhesive friction (EMC) are presented and compared with each other.

\subsubsection{Test 1: Mode II debonding}
\begin{figure}[h]
	\centering
	\includegraphics[height=0.35\linewidth]{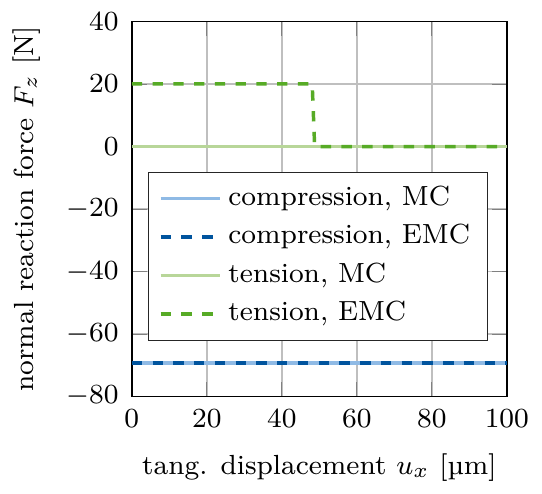}
		\includegraphics[height=0.35\linewidth]{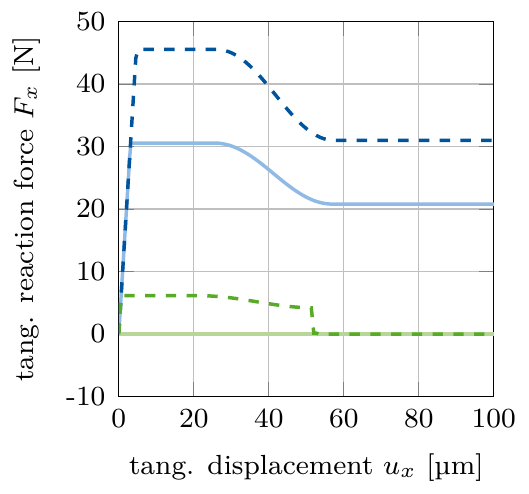}
	\caption[]{CSI debonding: Variation of the normal reaction $F_z$ (left) and the tangential reaction force $F_x$ (right) as a function of the tangential displacement $u_x$ for mode II debonding under constant compression (mode IIa) or constant tension (mode IIb). The results illustrate the difference between the modified Coulomb's law (MC) and the new extension (EMC).}
	\label{img:csi2_sliding} 
\end{figure}

Figure~\ref{img:csi2_sliding} shows the normal and tangential reaction forces $F_z$ and $F_x$ for debonding and possible subsequent sliding in (tangential) $x$-direction under prescribed constant compression or tension.
For a constant compression of -70~N, the slope of the curve of the tangential reaction force is identical for the MC and the EMC.
The maximum tangential reaction force increases from 30~N to 45~N for the EMC. 
For a constant tension of 20~N, the tangential reaction force reaches up to 6 N before decreasing and dropping to 0 because of the absence of contact.
The maximum tangential reaction force under tension is smaller than under compression, due to the decrease in $\phi$ stemming from the accumulated deformation in normal direction before the debonding started (due to pulling the implant back up before sliding).
The contact is lost abruptly after the limit for the accumulated deformation $g_\text{d}$ is reached, due to the positive contact gap at the bone-implant interface. 

\subsubsection{Test 2: Mode III debonding}
\begin{figure}[H]
	\centering
	\includegraphics[height=0.35\linewidth]{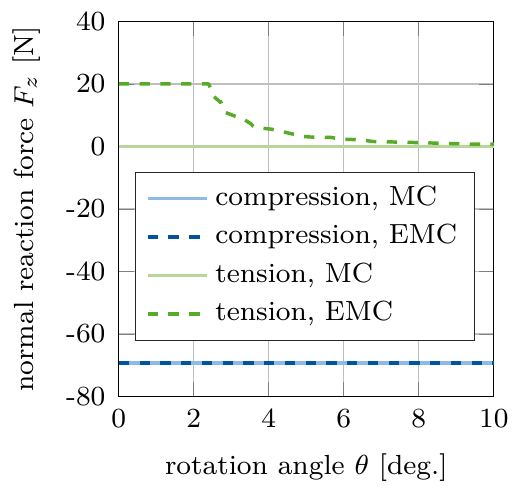}
	\includegraphics[height=0.35\linewidth]{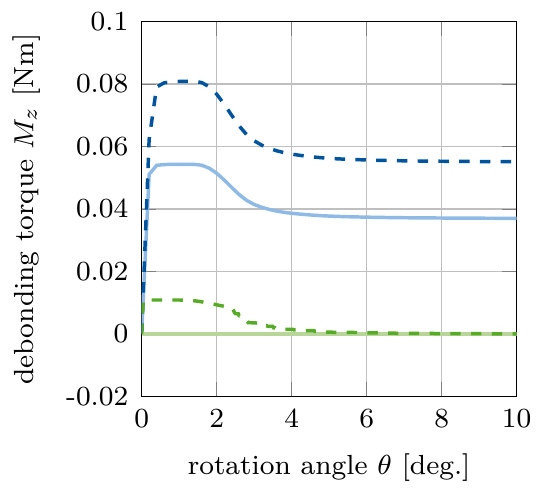}
	\caption[]{CSI debonding: Variation of the normal reaction $F_z$ (left) and the torque $M_z$ (right) as a function of the rotation angle $\theta$ for mode III debonding under constant compression (IIIa) or constant tension (IIIb). The results illustrate the difference between the modified Coulomb's law (MC) and its new extension (EMC).}
	\label{img:csi2_srotation} 
\end{figure}

Figure~\ref{img:csi2_srotation} shows the normal reaction force $F_z$ and the debonding torque $M_z$ for mode III debonding due to rotation around the implant's (normal) $z-$axis under prescribed constant compression or tension for the considered contact laws.
For a constant compression of -70~N, the slope of the torque curve is identical for both contact laws.
The maximal torque increases by 0.027 Nm (about 50\%) when including normal adhesion (EMC).
For a constant tension of 20~N the torque reaches 0.011~Nm and then decreases down to zero due to loss of contact.
This loss is gradual, starting in the external region of the cylinder and propagating inward to its center.
These results emphasize the fact that torque tests yield a stable crack propagation, which is particularly interesting when it comes to assessing the effective adhesion energy of the bone-implant interface~\citep{mathieu2012}.

\subsubsection{Test 3: Mode I+II debonding for varying angles}
\begin{figure}[H]
	\centering
	\includegraphics[height=0.35\linewidth]{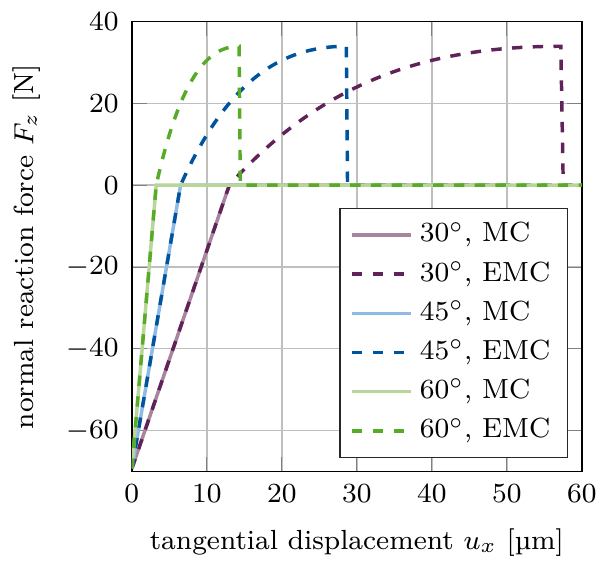}
	\includegraphics[height=0.35\linewidth]{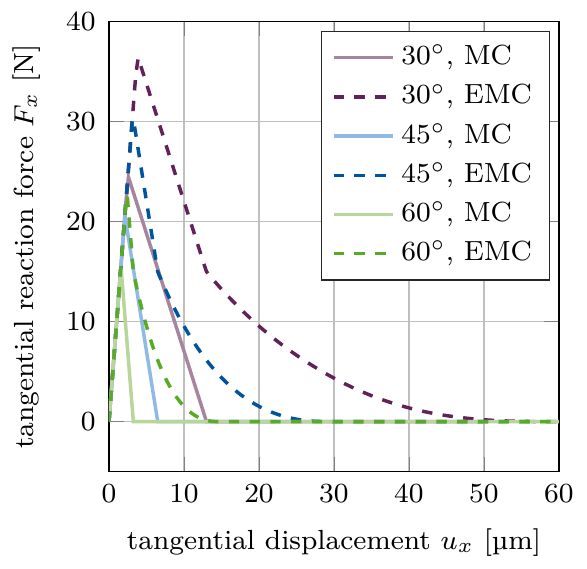}
	\caption[]{CSI debonding: Variation of the normal reaction force $F_z$ (left) and the tangential reaction force $F_x$ (right) as a function of the tangential displacement $u_x$ for mixed mode debonding, starting from an initial contact pressure (mode I+IIci). The results illustrate the difference between the modified Coulomb's law (MC) and its new extension (EMC).}
	\label{img:ACI2_CSI_ang}
\end{figure}

Figure~\ref{img:ACI2_CSI_ang} shows the normal and tangential reaction forces $F_z$ and $F_x$ for mixed mode debonding under different angles $\alpha$ (mode I+IIci) starting from an initial contact pressure, based on the MC and the EMC.
The normal reaction force $F_z$ increases linearly until it reaches zero. 
For each angle, the slope of the reaction force curve is identical for both considered contact laws, respectively.
In case of the EMC, the reaction force becomes positive at some point and follows the debonding curve of cohesive zone model \eqref{eq:adhesion_traction} seen in Fig.~\ref{img:normal_traction}.
In all presented cases, the debonding occurs because the maximal normal gap exceeds $g_\text{b}$, due to the prescribed upward movement.
Therefore, increasing the debonding angle $\alpha$ decreases the amount of tangential deformation necessary for debonding, i.e. where contact is lost and the reaction force becomes zero.

\begin{figure}[H]
	\centering
	\includegraphics[height=0.35\linewidth]{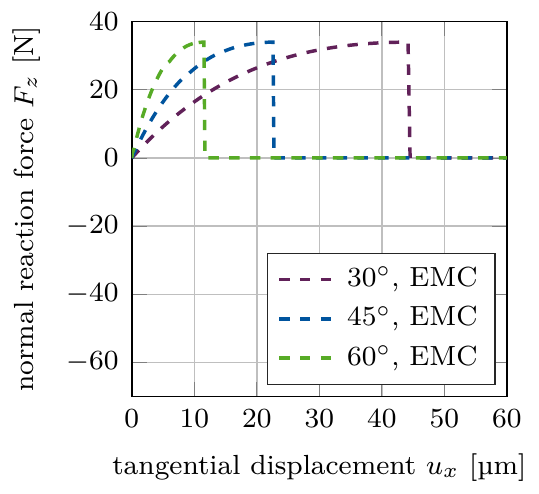}
	\includegraphics[height=0.35\linewidth]{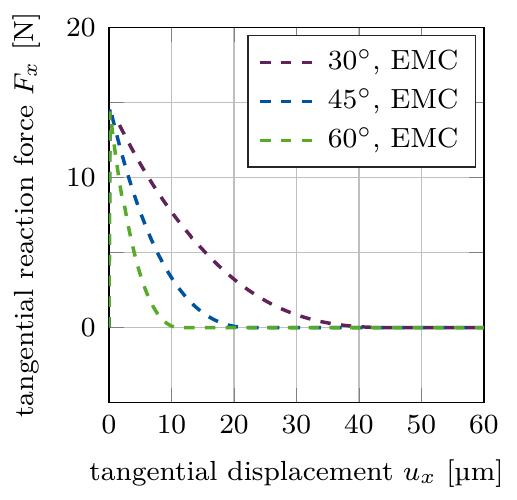}
	\caption[]{CSI debonding: Variation of the normal reaction force $F_z$ (left) and the tangential reaction force $F_x$ (right) as a function of the tangential displacement $u_x$ for mixed mode debonding, starting from zero contact force (mode I+IIdi). The results show the extended modified Coulomb's law (EMC).}
	\label{img:ACI2_CSI_ang2}
\end{figure}

The tangential reaction force $F_x$ increases linearly until the respective sliding limit is reached.
Then the implant starts sliding and the tangential reaction force decreases linearly, as long as the normal force $F_z$ is still negative. 
When the normal reaction force reaches zero, the cases with the MC show zero tangential reaction force, as there is no contact anymore.
For the cases with the EMC, the bone-implant interface has not fully debonded yet and thus, there is still a normal (adhesive) contact force building up. 
As a result, the tangential reaction force decreases nonlinearly until it reaches zero. While the maximum normal reaction force is the same for all three tested angles, the maximum tangential reaction force decreases for an increasing debonding angle.
The respective maximal tangential reaction force for each case is around 1.5 times higher for the EMC compared to the MC, due to the shift in tangential contact
traction~\eqref{eq:adhesive_friction}.
Figure~\ref{img:ACI2_CSI_ang2} shows the normal and tangential reaction forces $F_z$ and $F_x$ for mixed mode debonding under different angles $\alpha$ (mode I+IIdi), with no initial contact force, based on the EMC.
The different curves of the normal reaction force $F_z$ are identical to the positive part of the curves in Figure~\ref{img:ACI2_CSI_ang}.
As there is no initial compression or tension in the beginning of this loading case, all curves begin at the origin.
Similarly, the different curves  for the tangential reaction force $F_x$ are identical to the exponential part of the curves in Figure~\ref{img:ACI2_CSI_ang} and are shifted by the same displacement towards the origin, respectively, as the normal reaction force.
\subsubsection{Test 4: Mode I+II debonding for varying CZM parameter $t_0$}
\begin{figure}[H]
	\centering
	\includegraphics[height=0.35\linewidth]{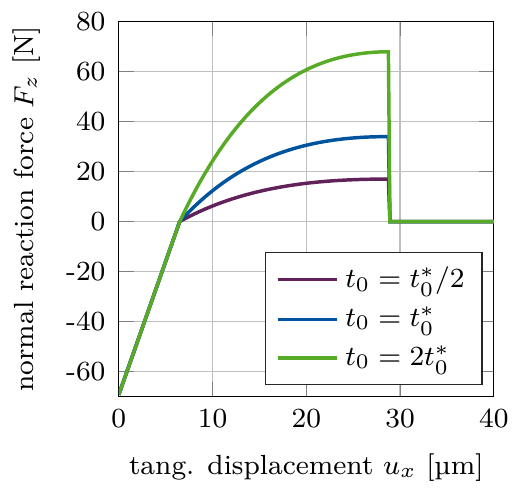}
	\includegraphics[height=0.35\linewidth]{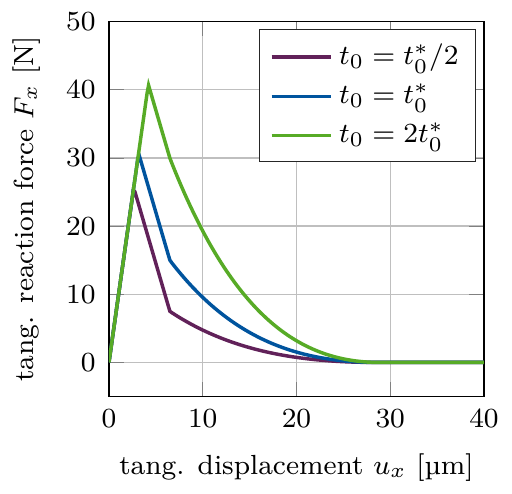}
	\caption[]{CSI debonding: Variation of the normal reaction force $F_z$ (left) and the tangential reaction force $F_x$ (right) as a function of the tangential displacement for mixed mode debonding under $\alpha=45^\circ$ and varying CZM parameter $t_0$, starting from an initial contact pressure (mode I+IIcii). The results show the extended modified Coulomb's law (EMC).}
	\label{img_csi2_ang_t0}
\end{figure}

\begin{figure}[H]
	\centering
	\includegraphics[height=0.35\linewidth]{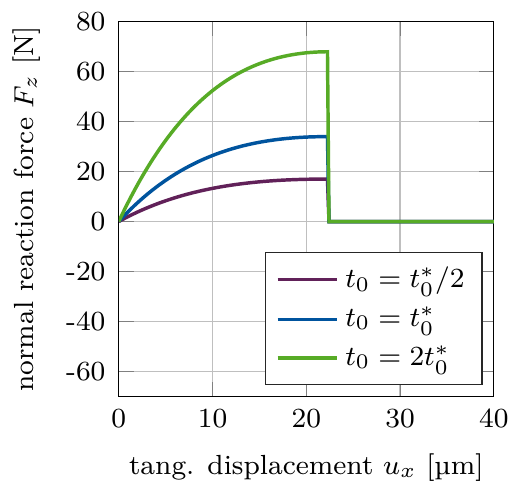}
	\includegraphics[height=0.35\linewidth]{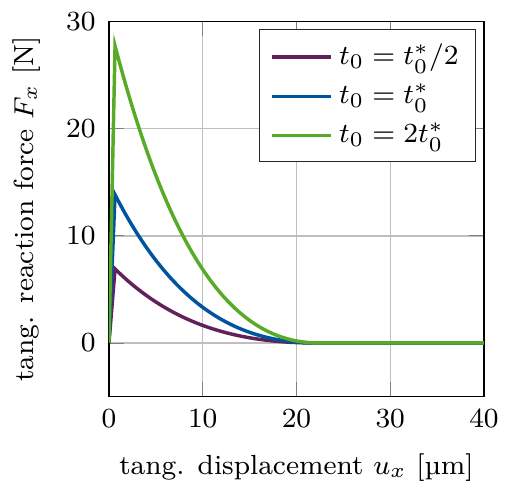}
	\caption[]{CSI debonding: Variation of the normal reaction force $F_z$ (left) and the tangential reaction force $F_x$ (right) as a function of the tangential displacement for mixed mode debonding under $\alpha=45^\circ$ and varying CZM parameter $t_0$, starting from zero contact force (mode I+IIdii). The results show the extended modified Coulomb's law (EMC).}
	\label{img_csi2_ang_t02}
\end{figure}

Figure~\ref{img_csi2_ang_t0} shows the normal and tangential reaction force $F_z$ and $F_x$ for mixed mode debonding under $\alpha=45^\circ$ as a function of the tangential displacement with different values of the maximal CZM traction $t_0$ when considering adhesive friction (mode I+IIcii). 
As expected, the maximum normal and tangential reaction forces are proportional to $t_0$ (see Eqs.~\eqref{eq:adhesion_traction} and \eqref{eq:adhesive_friction}).
Furthermore, the slopes of $F_z$ and $F_x$ at the transition from compression to tension ($F_x=0$, $u_x=6.5$ mm) are smooth for about $t_0=2t^*_0$ (as $2t^*_0 \approx \epsilon_\text{n}g_0/\phi_0 e$ (see Eq.~\ref{eq:smooth})). 
Figure~\ref{img_csi2_ang_t02} shows the normal and tangential reaction forces $F_z$ and $F_x$ for mixed mode debonding under $\alpha=45^\circ$, with no initial contact force, as a function of the tangential displacement for different values of the maximal CZM traction $t_0$ when considering adhesive friction (mode I+IIdii).
The different curves of the normal reaction force $F_z$ are identical to the positive part of the curves in Figure~\ref{img_csi2_ang_t0}.
As there is no initial contact force in the beginning of this loading case, all curves begin at the origin.
Similarly, the different curves of the tangential reaction force $F_x$ are identical to the exponential part of the curves in Figure~\ref{img_csi2_ang_t0} and are shifted by the same displacement towards the origin, respectively, as the normal reaction force.
\subsubsection{Test 5: Mode I+II debonding for varying degree of osseointegration}
\begin{figure}[H]
	\centering
	\includegraphics[height=0.35\linewidth]{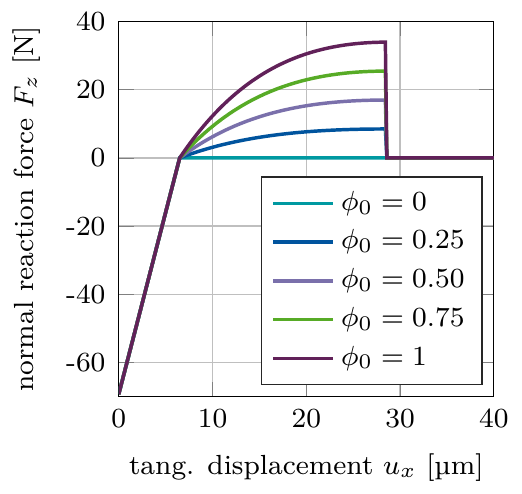}
	\includegraphics[height=0.35\linewidth]{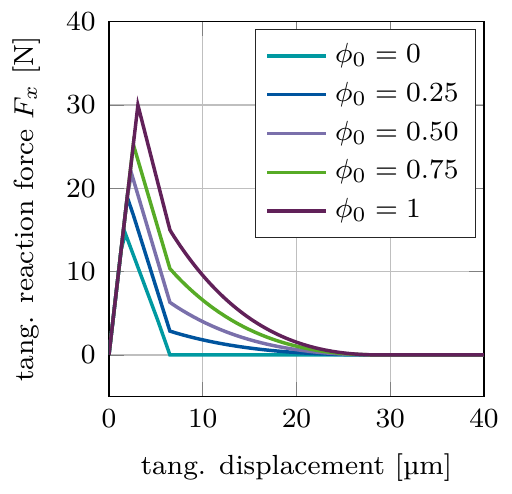}
	\caption[]{CSI debonding: Variation of the normal reaction force $F_z$ (left) and the tangential reaction force $F_x$ (right) as a function of the tangential displacement for mixed mode debonding under $\alpha=45^\circ$ and initial degree of osseointegration $\phi_0$,  starting from an initial contact pressure (mode I+IIciii). The results show the extended modified Coulomb's law (EMC).}
	\label{img:csi2_ang_phi}
\end{figure}

\begin{figure}[H]
	\centering
	\includegraphics[height=0.35\linewidth]{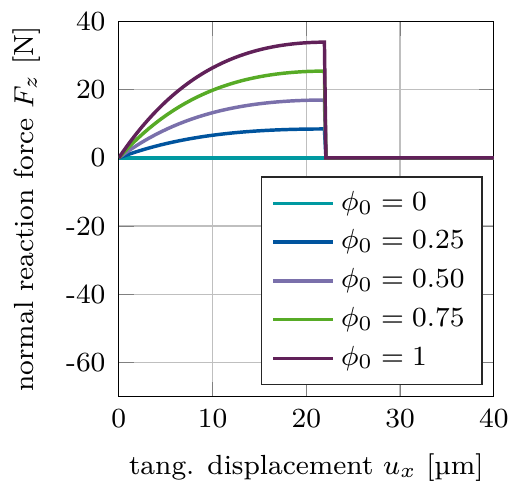}
	\includegraphics[height=0.35\linewidth]{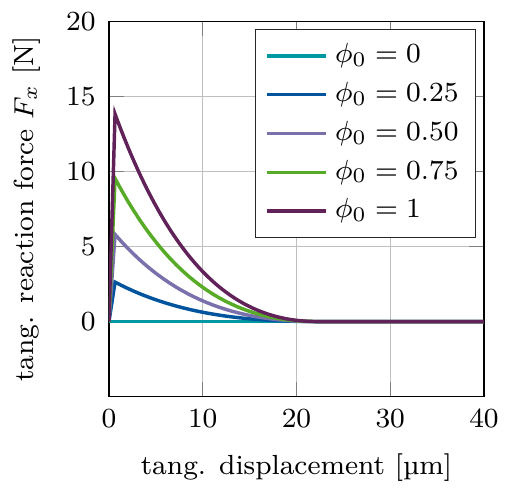}
	\caption[]{CSI debonding: Variation of the normal reaction force $F_z$ (left) and the tangential reaction force $F_x$ (right) as a function of the tangential displacement for mixed mode debonding under $\alpha=45^\circ$ and initial degree of osseointegration $\phi_0$,  starting from zero contact force (mode I+IIdiii). The results show the extended modified Coulomb's law (EMC).}
	\label{img:csi2_ang_phi2}
\end{figure}

Figure~\ref{img:csi2_ang_phi} shows the normal and tangential reaction force $F_z$ and $F_x$ for mixed mode debonding under 45$^\circ$ as a function of the tangential displacement with increasing degree of osseointegration $\phi_0$ when considering adhesive friction (mode I+IIciii).
Increasing the degree of initial osseointegration increases the peak magnitude in the normal and tangential reaction force.
This is due to the fact that in this test case, debonding occurs first due to $g_\text{n} > g_\text{b}$ and not due to exceeding the limit of the deformation of the interface $g_\text{d} = a_\text{s}(1+2b_\text{s})$ (see Eq.~\eqref{eq:phi}).
The maximal normal and tangential reaction force increase proportionally with increasing $\phi_0$, while the tangential displacement necessary for debonding remains the same.
Figure~\ref{img:csi2_ang_phi2} shows the normal and tangential reaction forces $F_z$ and $F_x$ for mixed mode debonding under $\alpha=45^\circ$, with no initial contact force, as a function of the tangential displacement with increasing degree of osseointegration $\phi_0$ when considering adhesive friction (mode I+IIdiii).
The different curves of the normal reaction force $F_z$ are identical to the positive part of the curves in Figure~\ref{img:csi2_ang_phi}.
As there is no initial contact force in the beginning of this loading case, all curves begin at the origin.
Similarly, the different curves of the tangential reaction force $F_x$ are identical to the exponential part of the curves in Figure~\ref{img:csi2_ang_phi} and are shifted by the same displacement towards the origin, respectively, as the normal reaction force.

%
\section{Application to acetabular cup implants} \label{s:ACI2_ACI2}
Both, the MC and the EMC have been examined for a simple implant model in Section~\ref{s:csi_validation}.
Now, both models are applied to simulate the debonding of a 3D, osseointegrated acetabular cup implant (ACI), under different removal conditions, similar to the simulations in~\cite{raffa2019} and \cite{immel2021}.
However, the aforementioned works only considered primary stability.
Here, the implant's secondary stability is considered and is quantified by assessing the removal force/torque.
Furthermore, the biomechanical determinants of the long-term stability, such as primary stability and initial degree of osseointegration are assessed.
The results of the modified Coulomb's law (MC) and its extension to adhesive friction (EMC) are compared to assess the importance of adhesive effects for long-term stability because it allows to distinguish the influence of primary stability and osseointegration phenomena on the secondary stability.
%
\subsection{Setup} \label{s:ACI2_setup2}
A simple cylindrical block is considered, as it is a suitable simplification of the pelvis geometry that qualitatively captures the relevant contact conditions.
The same geometry of the ACI including the ancillary used in ~\cite{raffa2019} and \cite{immel2021} is considered herein and is briefly summarized in the following.
An idealized bone block with the same properties as in~\cite{raffa2019} is used in order to calibrate the model and compare results.
The bone block is modeled as a cylinder with a radius of 50 mm and a height of 40 mm.
A hemi-spherical cavity is cut into the cylinder with a radius $R_\text{b}$ based on the fixed radius of the implant $R_\text{i}$ and the chosen interference fit $I\!F$, i.e., $R_\text{b}=R_\text{i}-I\!F/2$.
The edge of the cavity has a fillet radius of 2 mm.

\begin{figure}[H] 
	\centering
			\includegraphics[width=0.25\linewidth]{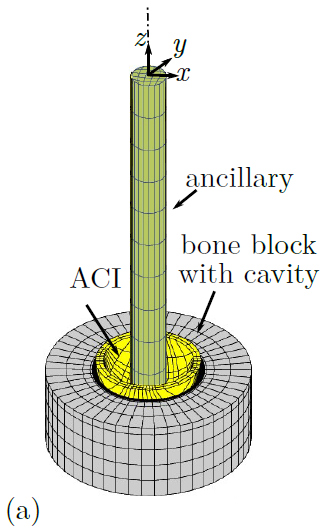}
	\hspace{5 mm}
	\begin{overpic}[scale=.3]
		{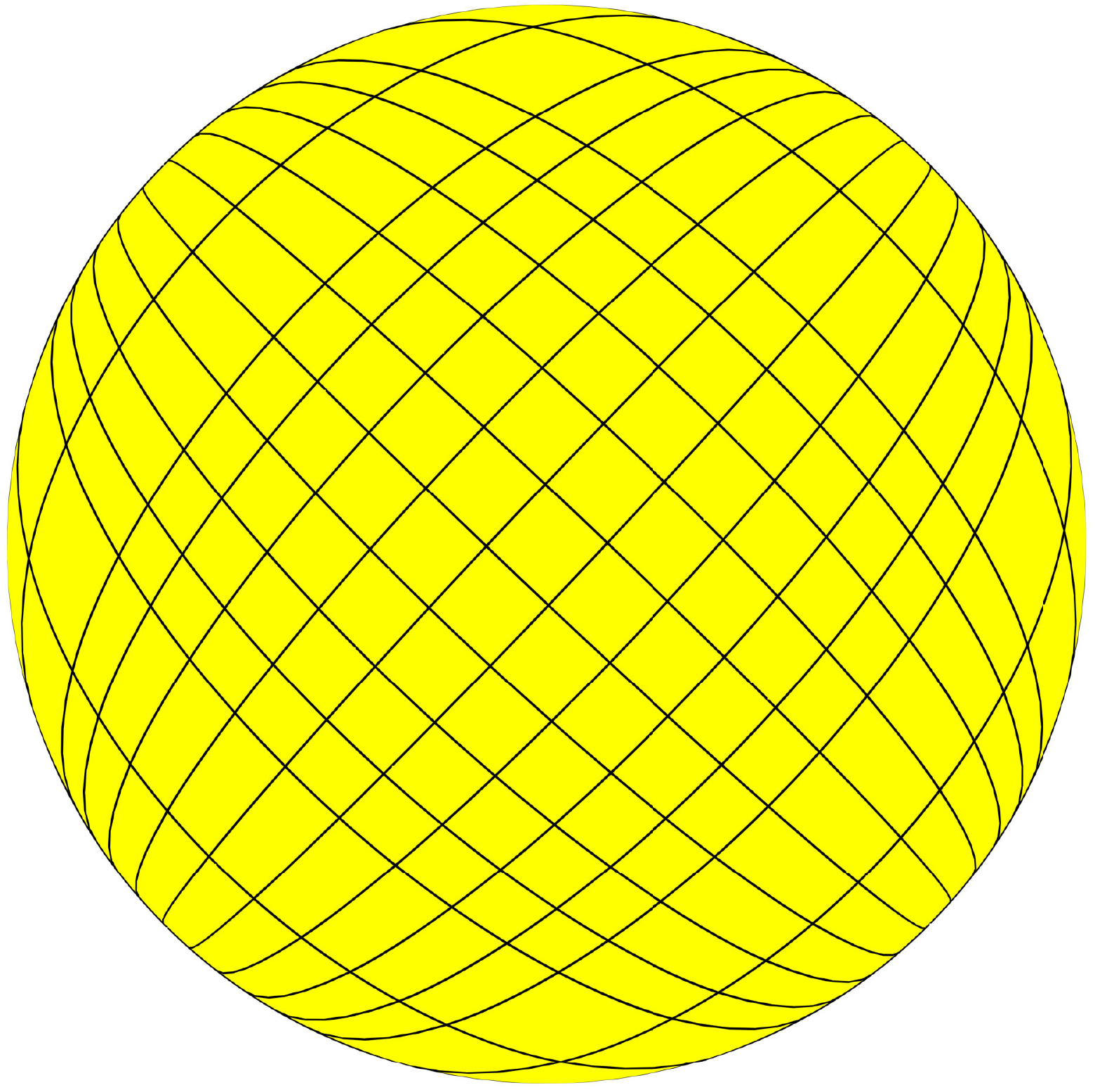}
		\put(0,0){(b)}
	\end{overpic}
	\hspace{5mm}
	\begin{overpic}[scale=.3]
		{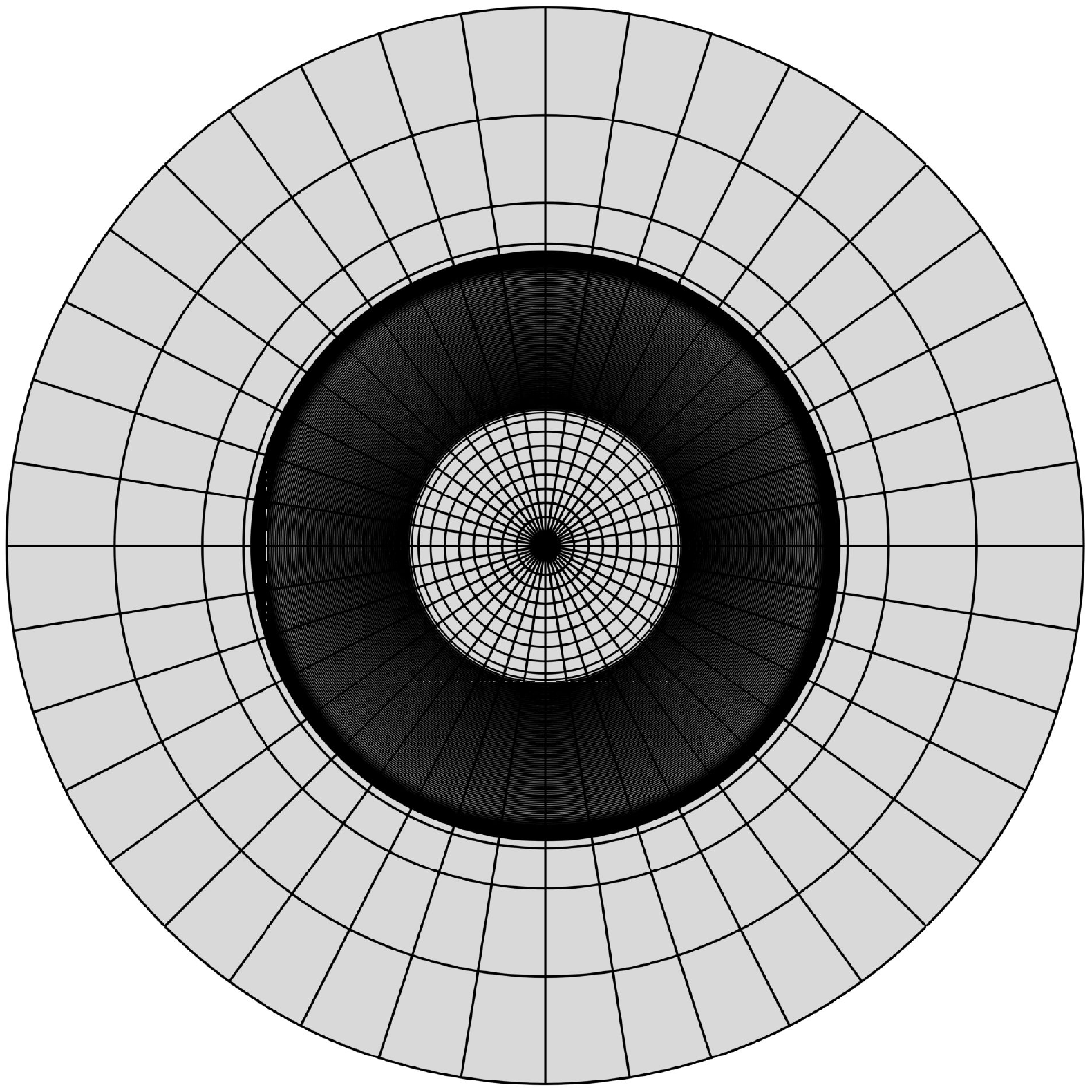}
		\put(0,0){(c)}
	\end{overpic}
	\caption[]{(a) Finite element mesh of the ACI, bone block, and ancillary~\citep{immel2021b}. (b) Bottom view of the ACI. (c) Top view of the bone block. A very fine mesh is used around the rim of the cavity, as the contact forces are expected to vary most strongly there.}
	\label{img:aci2_mesh}
\end{figure}

As in Section~\ref{s:csi_validation}, the bodies are meshed with surface-enriched hexahedral elements according to the parameters given in Table~\ref{tab:mesh}.
The finite element mesh is shown in Figure~\ref{img:aci2_mesh}. 
A refinement analysis of the mesh and the load step size is performed to ascertain mesh convergence (see Appendix A) for the reference case (see Section~\ref{s:aci2_setup_params}). 

\begin{table}[H]	
	\centering
	\begin{tabular}{lccc}
		\hline 
		body & $n_\text{el}$ & shape fcts. & $n_\text{gp}$\\ 
		\hline 
		implant bulk & 338  & lin. Lagrange & $2\times 2\times 2$ \\
		ancillary & 250 & lin. Lagrange & $2\times 2\times 2$  \\
		bone bulk & 20000 & lin. Lagrange & $2\times 2\times 2$ \\
		outer implant surf. & 169 & quad. NURBS & $5\times 5$ \\
		upper bone surf. & 4000  & quad. NURBS & $5\times 5$\\
		\hline
	\end{tabular} 
	\caption[]{ACI debonding: Number of finite elements $n_\text{el}$, type of shape functions and number of Gauss points per element $n_\text{gp}$ for the two bodies and their contact surfaces.}
	\label{tab:mesh}
\end{table}
%
\subsubsection{Model parameters} \label{s:aci2_setup_params}
The ancillary and the ACI are assumed to be made of stainless steel ($E_\text{a}=211$ GPa, $\nu_\text{a}=0.3$) and titanium alloy (Ti-Al6-V4; $E_\text{i}=113$ GPa, $\nu_\text{i}=0.3$), respectively.
The bone block is assumed to consist only of trabecular bone tissue ($\nu_\text{b}=0.3$~\citep{yew2006}), without an outer cortical layer.
This is in accordance with a previous study~\citep{immel2021} and findings in the literature~\citep{anderson2005,phillips2007,watson2017}, that indicate that the reaming of the hip performed during surgery may completely remove cortical bone tissue from the contact area.
For all bodies, the Neo-Hookean material model of Eq.~\eqref{eq:neohooke} is used. 
Furthermore, all materials are assumed to be homogeneous and isotropic and both contact surfaces are assumed to be perfectly smooth.

In this work, the effect of various biomechanical properties of the bone-implant system on the ACI long-term stability is assessed.
Therefore, different degrees of osseointegration from $0-100$\% with homogeneous distribution over the contact surfaces are considered.
Furthermore, the influence of varying bone stiffness $E_\text{b}=0.1-0.6$ GPa (\cite{phillips2007,janssen2010}), interference fit $I\!F=0-2.0$ mm (\cite{kwong1994,macdonald1999, spears1999}), and friction coefficient $\mu_\text{b}=0-0.7$ (\cite{dammak1997a,spears1999,novitskaya2011}) on the long-term stability are analyzed.
The corresponding friction coefficient $\mu_\text{ub}=0.15-1$ is taken from Table 4 from~\cite{immel2020} and is roughly 1.5 times higher than $\mu_\text{b}$.
Based on previous studies~\citep{raffa2019,immel2021} the parameter set of $E^*_\text{b}= 0.2$ GPa (\cite{phillips2007}), $I\!F^*=1$ mm (\cite{kwong1994}), and $\mu^*_\text{b}=0.3$ (\cite{dammak1997a}) is denoted as the reference case and marked with $\ast$.
The parameters of the state function~\eqref{eq:phi} are chosen to be $a_\text{s}=128$ µm and $b_\text{s}=1.84$, which does not affect the maximum of the removal force/torque.
Due to the lack of experimental data, the values of $a_\text{s}$ and $b_\text{s}$ are chosen large enough so that the debonding process is visible and a removal force/torque can be identified (Fig.~\ref{img:appndx_a_and_b}).
The coefficients of the cohesive zone model $t_0=t^*_0=1.8$ MPa and $g_\text{b}=g_0=3a_\text{s}$ are calibrated based on the results of~\cite{ronold2002} for polished CSI, as was done in Section~\ref{s:csi_setup}.
All parameters of the contact model and their values, as well as the studied parameters are listed in Table~\ref{tab:aci_params}.
\begin{table}[H]
	\centering
	\begin{tabular}{ccc}
		\hline 
		param. & value & source \\ 
		\hline 
		$a_\text{s}$ & 128 µm & chosen reasonably \\
		$b_\text{s}$ & 1.84 & chosen reasonably \\
		$\mu_\text{ub}$ & 1.5$\mu_\text{b}$ &  \cite{immel2020} \\
		$t_0$ & 1.8 MPa & \cite{ronold2002} \\
		$g_0$ & 3$a_\text{s}$ & chosen reasonably\\
		$g_\text{b}$ & $3a_\text{s}$ & chosen reasonably\\
		\hline
		$\phi_0$ & 0 - 1 & chosen reasonably \\
		$\mu_\text{b}$ & 0 - 0.7 & $[$\citeyear{dammak1997a,spears1999,novitskaya2011}$]$\\
		$E_\text{b}$ & 0.1 - 0.6 GPa & $[$\citeyear{phillips2007,janssen2010}$]$ \\
		$I\!F$ & 0 - 2 mm & $[$\citeyear{kwong1994,macdonald1999, spears1999}$]$\\
		\hline
	\end{tabular}
	\caption[]{List of the constant parameters of the modified Coulomb friction law and the chosen values, as well as the range for the varied parameters.}
	\label{tab:aci_params}
\end{table}

%
\subsubsection{Boundary and loading conditions and solver settings} \label{s:setup_bc}
\begin{figure}[H]
	\centering
	\begin{subfigure}[t]{0.6\linewidth}
		\centering
		\includegraphics[width=\linewidth]{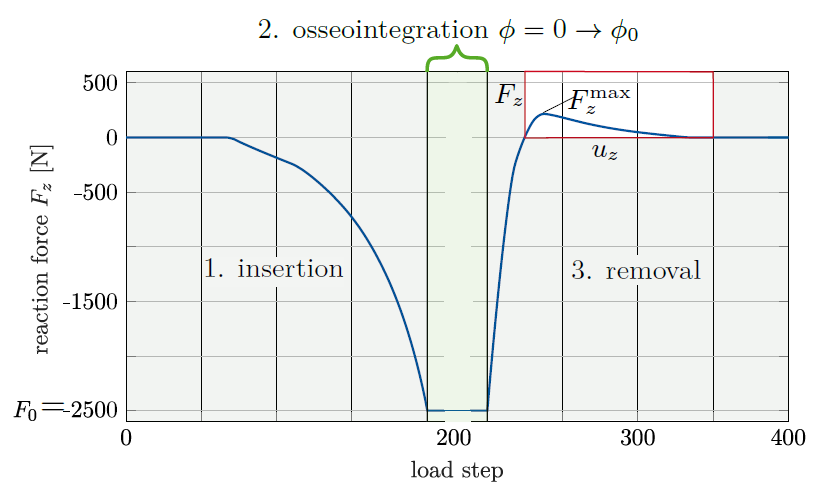}
		\caption[]{Illustration of the three simulation stages on the example of mode I debonding. The final output of the reaction force and maximum pull-out force is shown in red. }
		\label{img:Ra}
	\end{subfigure}
	\hfill
	\begin{subfigure}[t]{0.37\linewidth}
		\centering
		\includegraphics[width=0.8\linewidth]{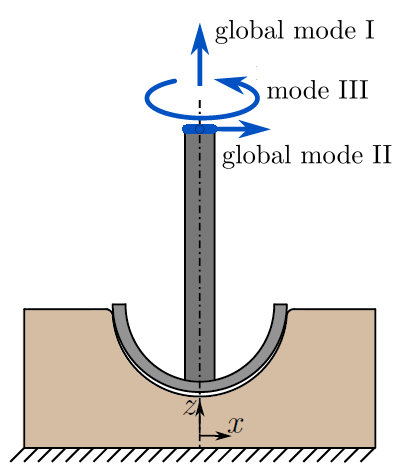}
		\caption{Illustration of the three different removal tests.}
		\label{img:aci2_bc}
	\end{subfigure}
	\caption[]{ACI debonding: (a) Illustration of the three simulation stages and (b) the three removal tests.}
	\label{img:cmt_setup}
\end{figure}

The bone block is fixed in all directions at the bottom surface. 	
As before, only quasi-static conditions are considered.
The simulations of implant insertion and subsequent removal are comprised of three stages (see Fig.~\ref{img:cmt_setup}):
\begin{enumerate}
	\item insertion: the implant is inserted vertically into the cavity, by pushing the upper surface of the ancillary in negative $z$-direction, until the reaction force at the top of the ancillary reaches $F_0=-2500$ N, similar to values found in the literature~\citep{sotto2010,souffrant2012,lecann2014} and to what was done in previous studies~\citep{raffa2019,immel2021}.
	The downward displacement attained at the top of the ancillary for $F_0=-2500$~N is denoted $d_0$.
	It depends on the considered parameters $E_\text{b}, I\!F$, and $\mu_\text{b}$ and thus changes for each case, i.e.~$d_0=d_0$($\mu_\text{b}$, $I\!F$, $E_\text{b}$, $F_0$).
	\item osseointegration: the contact surfaces are assumed to be homogeneously osseointegrated  with an initial degree of osseointegration varying from $\phi_0\in[0, 0.25, 0.5, 0.75, 1]$.
	\item removal: the implant is removed either
	\begin{itemize}
		\item global\footnote{which induces local mixed mode I + II debonding} mode I: by displacing the upper surface of the ancillary in positive $z$-direction by -$d_0$,
		\item global mode II: by displacing the center of the upper surface of the ancillary in positive $x$-direction by $d_0$, 
		\item mode III: by rotating the upper surface of the ancillary around its $z$-axis by $\theta=10^\circ$.
	\end{itemize} 
\end{enumerate}

The three simulation stages are shown in Figure~\ref{img:Ra} and the different removal cases are illustrated in Figure~\ref{img:aci2_bc}.
The example of mode I debonding is shown, with the final output of the load-displacement curve inside the red square (Fig.~\ref{img:Ral}(a)).

The stability of the configuration is then assessed by determining the maximum pull-out force in normal direction, $F^\text{max}_z$, the maximum pull-out force in tangential direction, $F^\text{max}_x$, or the maximum debonding torque $M^\text{max}_\text{z}$.

Contact is computed with a penalty regularization, and the corresponding penalty parameter is chosen based on the Young's modulus of trabecular bone as $\epsilon_\text{n} = \epsilon_\text{t} = E_\text{b}/L_0$, with $L_0=R_\text{I}=0.0255$ m corresponding to the radius of the implant.
The number of load steps for the different simulation stages are: $ l_1=l_{3\text{.modeI}}=l_{3.\text{modeIII}}=100$ and $l_{3.\text{modeII}}=1000$.
All simulations were performed with an in-house, MATLAB-based solver (R2019b, The MathWorks, Natick, MA, USA) with MATLAB's own parallelization. 
Computations were performed on the RWTH Compute Cluster (Intel HNS2600BPB, Platinum 8160) with 20 cores.
The average computing time for the different contact laws and loading cases is listed in Table~\ref{tab:computing_time}.
The computing time is sensitive to the parameter combination. 
Parameter combinations that produce high pull-out forces/debonding torque have a longer computing time.
The difference in computing time between the debonding tests and the contact models is discussed in Section~\ref{s:convergence}.

%
\subsection{Debonding without adhesion in normal direction} \label{s:results}
First, the results of the removal tests, in terms of load-displacement curves and pull-out force/ debonding torque, obtained with the MC are presented. 
The results with the EMC follow in Section~\ref{s:results_adhesion}, and a comparison is given in Section~\ref{s:comparison}.
%
\subsubsection{Global mode I: Normal pull-out test} \label{s:results_normal} 
\begin{figure}[H]
	\centering
	\includegraphics[height=0.35\linewidth]{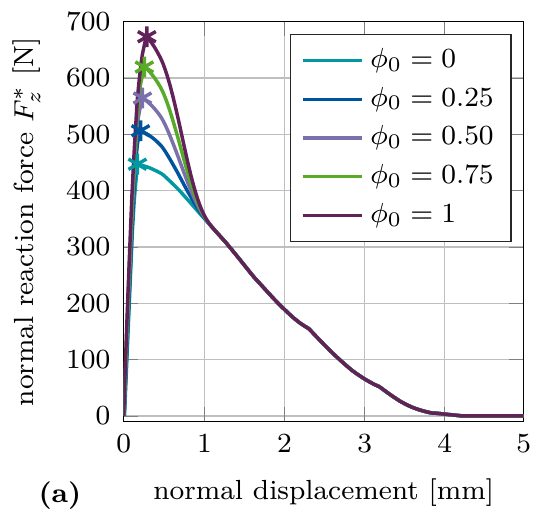}
	
	\includegraphics[height=0.35\linewidth]{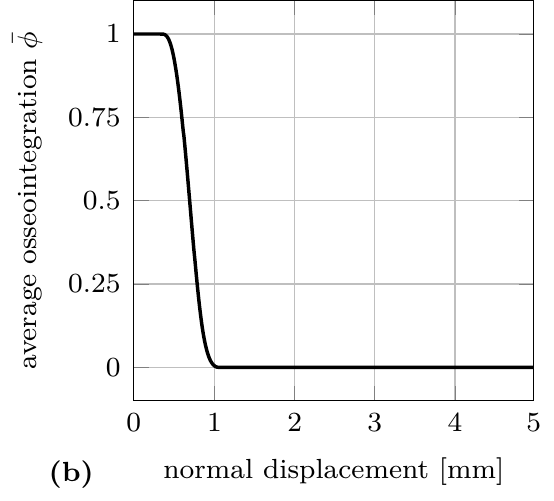}
	\caption[]{Normal debonding without adhesion for the reference case: (a) Variation of the normal force $F^*_z$ a function of the initial degree of osseointegration $\phi_0$. The maximum pull-out force $F^{\text{max}*}_z$ is marked with $\ast$. (b) Average degree of osseointegration of the bone-implant interface $\bar{\phi}$ for an initial degree of osseointegration $\phi_0=1$. }
	\label{img:PO_pull}
\end{figure}

\begin{figure}[H]
	\centering
	\includegraphics[width=0.5\linewidth]{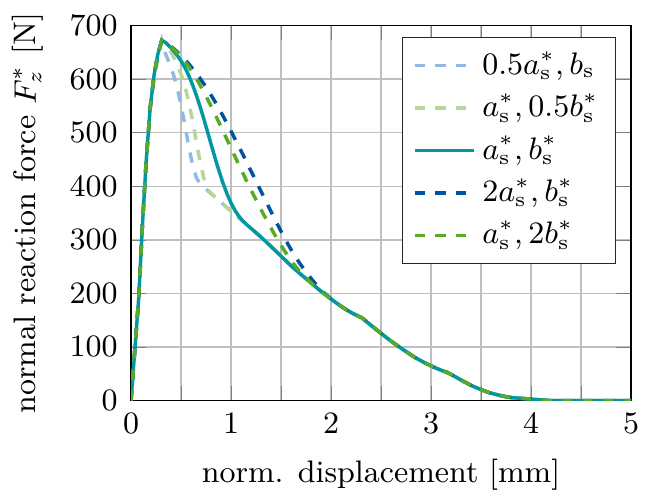}
	\caption[]{Normal debonding without adhesion: Variation of the normal force $F^*_z$ for $\phi_0=1$ as a function of the MC parameters $a_\text{s}$ and $b_\text{s}$,}
	\label{img:appndx_a_and_b}
\end{figure}

Figure~\ref{img:PO_pull} (a) shows the normal reaction force $F^*_z$ for the reference case, which increases and reaches a peak at a displacement of 0.25--0.32 mm and then slowly decreases to zero.
This maximum coincides with the start of the decrease of the average degree of osseointegration of the bone-implant interface $\bar{\phi}$ (Fig.~\ref{img:PO_pull}, (b)).
At a displacement of 1.07 mm, the reaction force becomes independent from $\phi_0$.
At this point, the bone-implant interface is completely debonded ($\bar{\phi}=0$) and only pure Coulomb's friction is taking place until the contact at the bone-implant interface is lost completely at a displacement of about 4.25 mm.
In this test, osseointegration only affects the magnitude of the peak, while the overall slope of the load-displacement curve remains unchanged when increasing  the initial degree of osseointegration.
The location of the peak does not change significantly with increasing $\phi_0$.
Due to the lack of experimental data for this work, the values of $a^*_\text{s}=128$ µm and $b^*_\text{s}=1.84$ are chosen large enough so that the debonding process is visible and a removal force/torque can be identified.
The effect of changing the value of $a_\text{s}$ and $b_\text{s}$ on $F^*_z(\phi_0=1)$ is shown in Figure~\ref{img:appndx_a_and_b}.
Naturally, both parameters have no effect on the mechanical behavior before debonding and on the maximum pull-out force.
Decreasing or increasing $a_\text{s}$ and $b_\text{s}$ decreases or increases the amount of deformation that is necessary for the interface to fully debond (about 0.7, 0.75, 1.1, 1.75, 1.9 mm, respectively, see Fig.~\ref{img:appndx_a_and_b}). 
After debonding, only pure Coulomb's friction takes place until the contact between bone and implant is lost completely (after a displacement of about 4.25 mm).

\begin{figure}[H]
	\centering
	\includegraphics[width=0.3\linewidth]{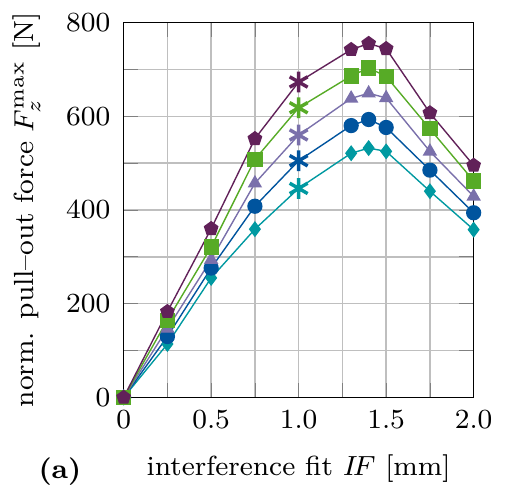}
	\hspace{-15mm}
	\includegraphics[width=0.54\linewidth]{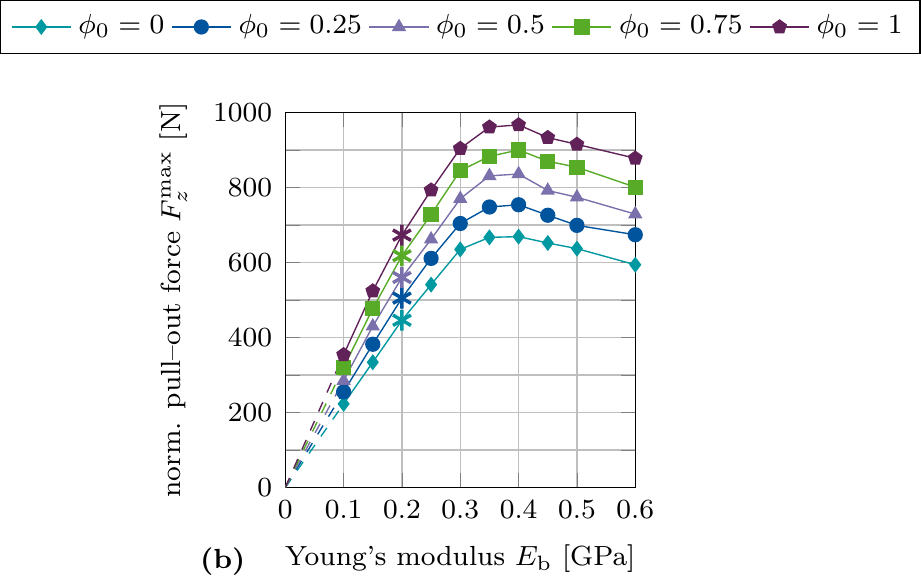}
	\hspace{-25mm}
	\includegraphics[width=0.305\linewidth]{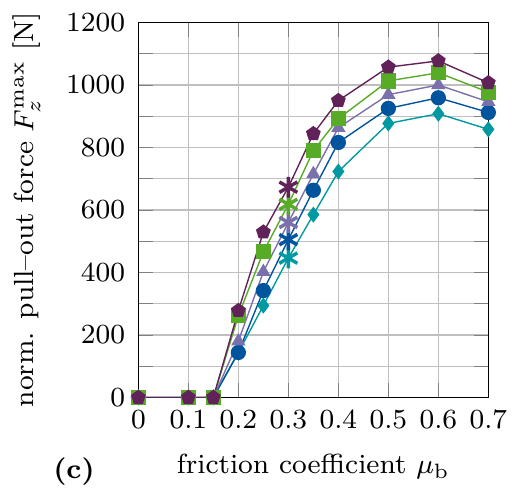}
	\caption[Normal debonding without adhesion.]{Normal debonding without adhesion: Variation of the maximum normal pull-out force $F^\text{max}_z$ as a function of the initial degree of osseointegration $\phi_0$ and (b) the interference fit $I\!F$, (c) the trabecular Young's modulus $E_\text{b}$, and (d) the friction coefficient $\mu_\text{b}$. The reference case is marked with $\ast$.}
	\label{img:PO_pull2}
\end{figure}

Figure \ref{img:PO_pull2}(a)--(c) shows the maximum normal pull-out force $F^\text{max}_z$ as a function of the interference fit $I\!F$, trabecular bone stiffness $E_\text{b}$, friction coefficient $\mu_\text{b}$, for different values of the initial degree of osseointegration $\phi_0$.
The results obtained for $F^\text{max}_z$ with $\phi_0=0$ are identical to the results from~\cite{raffa2019}.
First, the pull-out force increases as a function of $I\!F, E_\text{b}, \mu_\text{b}$, then reaches a peak, and eventually decreases.
The maximum value of the pull-out force is obtained for around $I\!F=1.4$ mm, $E_\text{b}=0.4$ GPa, and $\mu_\text{b}=0.6$.
For $\mu_\text{b} \le 0.15$ the pull-out force is zero, for all degrees of osseointegration.



\subsubsection{Global mode II: Tangential pull-out test}
\begin{figure}[H]
	\centering
	\includegraphics[height=0.35\linewidth]{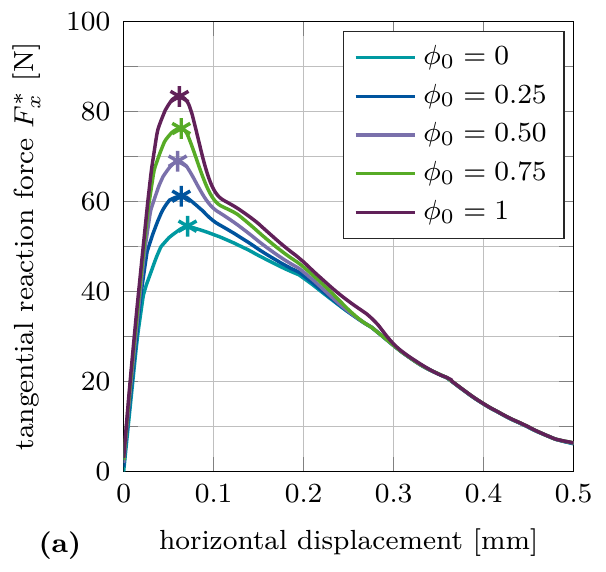}
	
	\includegraphics[height=0.35\linewidth]{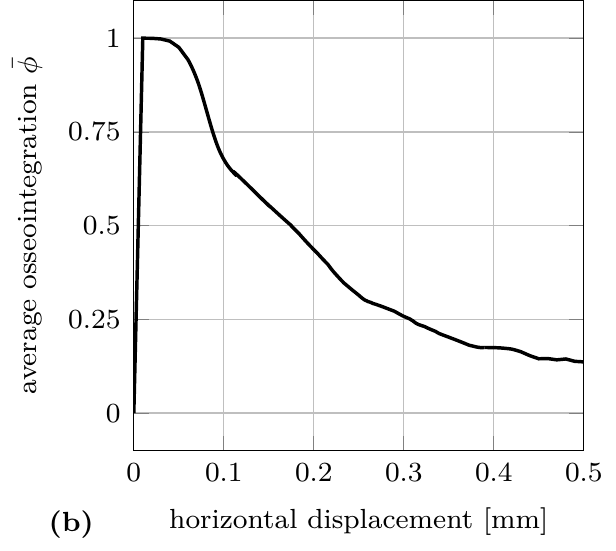}
	\caption[Tangential debonding without adhesion.]{Tangential debonding without adhesion for the reference case: (a) Variation of the tangential force $F^*_x$ as a function of the initial degree of osseointegration $\phi_0$. The maximum pull-out force $F^{\text{max}*}_x$ is marked with $\ast$. (b) Average degree of osseointegration of the bone-implant interface $\bar{\phi}$ for an initial degree of osseointegration $\phi_0=1$.}
	\label{img:PO_tang}
\end{figure}

The tangential reaction force $F^*_x$ for the reference case increases and reaches a peak at a displacement of about 75 µm and then slowly decreases to zero (Fig.~\ref{img:PO_tang}(a)).
The average degree of osseointegration of the bone-implant interface $\bar{\phi}$ starts to decrease already beyond 34 µm (Fig.~\ref{img:PO_tang} (b)). At a displacement of about 0.3 mm, the reaction force becomes independent from $\phi_0$.
Similarly to the normal pull-out test, increased osseointegration only affects the magnitude of the tangential pull-out force, while the location of the peaks and the initial slope of the curves for different degrees of osseointegration are very similar.
\begin{figure}[H]
	\centering
	\includegraphics[width=0.3\linewidth]{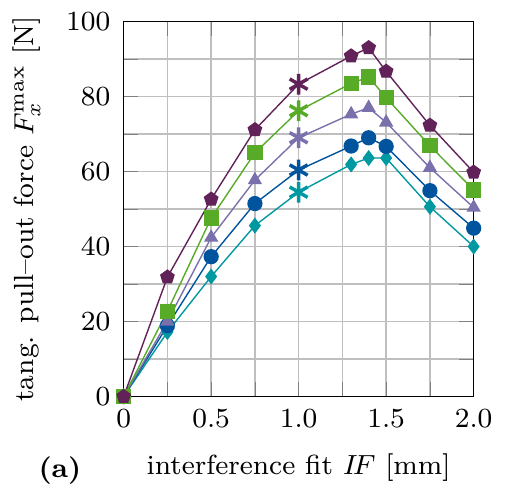}
	\hspace{-15mm}
	\includegraphics[width=0.545\linewidth]{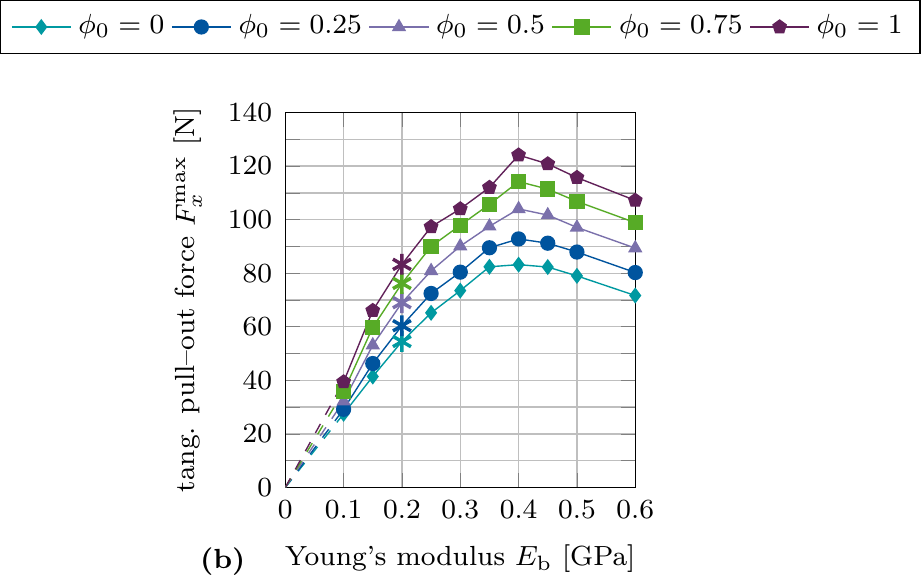}
	\hspace{-25mm}
	\includegraphics[width=0.3\linewidth]{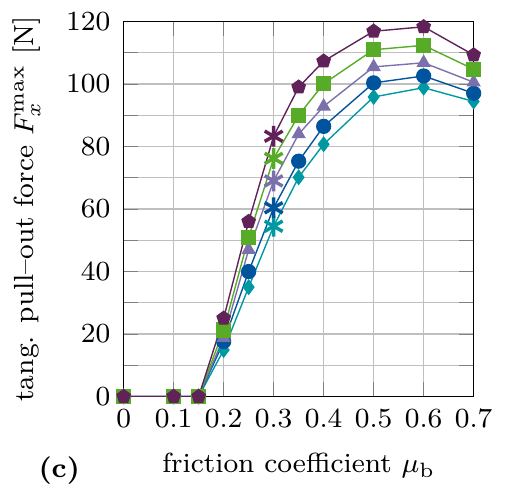}
	\caption[Tangential debonding without adhesion.]{Tangential debonding without adhesion: Variation of the maximum tangential pull-out force $F^\text{max}_x$ as a function of the initial degree of osseointegration $\phi_0$ and (a) the interference fit $I\!F$, (b) the trabecular Young's modulus $E_\text{b}$, and (c) the friction coefficient $\mu_\text{b}$. The reference case is marked with $\ast$.}
	\label{img:PO_tang2}
\end{figure}
Figure~\ref{img:PO_tang2}(a)--(c) shows the maximum tangential pull-out force $F^\text{max}_x$ as a function of the interference fit $I\!F$, trabecular bone stiffness $E_\text{b}$, friction coefficient $\mu_\text{b}$, for different values of the initial degree of osseointegration $\phi_0$.
First, the pull-out force increases as a function of $I\!F, E_\text{b}$ and $\mu_\text{b}$, then reaches a peak, and eventually decreases.
The maximum value of the pull-out force is obtained for around $I\!F=1.4$ mm, $E_\text{b}=0.4$ GPa, and $\mu_\text{b}=0.6$ -- the same values as for the normal pull-out test.
For $\mu_\text{b} \le 0.15$ the pull-out force is zero for all degrees of osseointegration.
Tangential pull-out forces are roughly one magnitude lower than the corresponding normal pull-out force, which agrees with observations from clinical practice.
During surgery, after the insertion of the ACI, surgeons often attempt to lever out an acetabular cup to test the seating of the ACI.
That is, the surgeon applies a tangential force, such as is considered here, instead of a normal force since normal pull-out would require too much force.
\subsubsection{Mode III: Torsional debonding test}
\begin{figure}[H]
	\centering
	\includegraphics[height=0.35\linewidth]{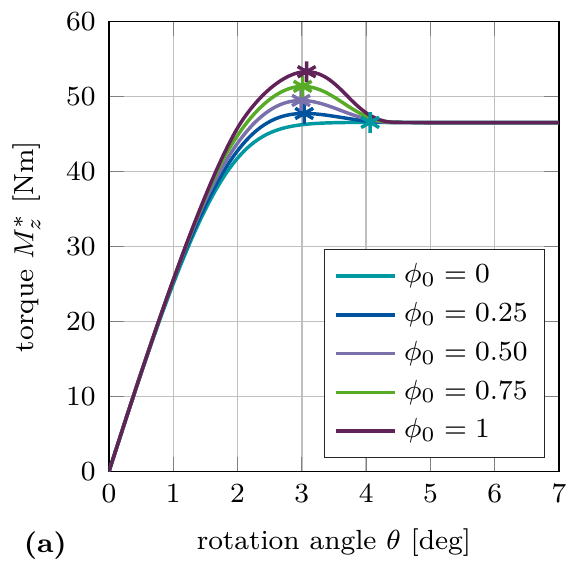}
	
	\includegraphics[height=0.35\linewidth]{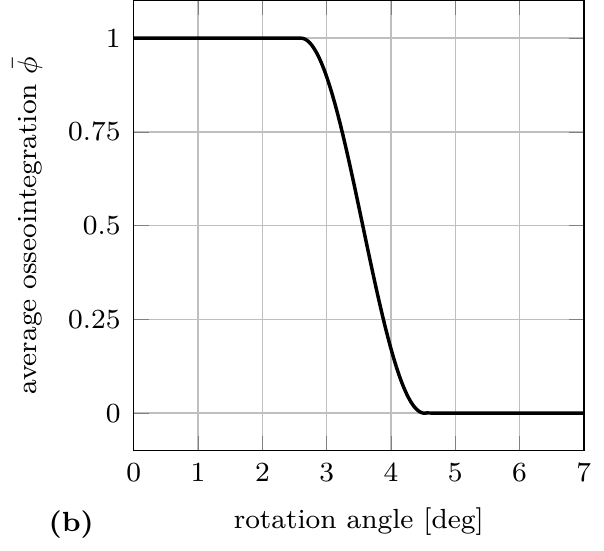}
	\caption[Torsional debonding without adhesion.]{Torsional debonding without adhesion for the reference case: (a) Variation of the debonding torque $M^*_z$ a function of the initial degree of osseointegration $\phi_0$. The maximum debonding torque $M^{\text{max}*}_z$ is marked with $\ast$. (b) Average degree of osseointegration of the bone-implant interface $\bar{\phi}$ for an initial degree of osseointegration $\phi_0=1$.}
	\label{img:PO_rot}
\end{figure}

Figure \ref{img:PO_rot} (a) shows the debonding torque $M^*_z$ as a function of the rotation angle for different values of $\phi_0$ and the reference case.
The torque increases, reaches a peak at an angle of about 3$^\circ$ and then decreases to reach a constant torque of 47 Nm at about 4.5$^\circ$ due to the present compressive normal force.
The degree of osseointegration starts to decrease at an angle of about 2.6$^\circ$ and becomes zero at about 4.5$^\circ$ (Fig.~\ref{img:PO_rot} (b)).
As for the normal and the tangential pull-out cases, only the magnitude of the peak of the load-displacement curve is affected when increasing the degree of osseointegration $\phi_0$. 

\begin{figure}[H]
	\centering
	\includegraphics[width=0.3\linewidth]{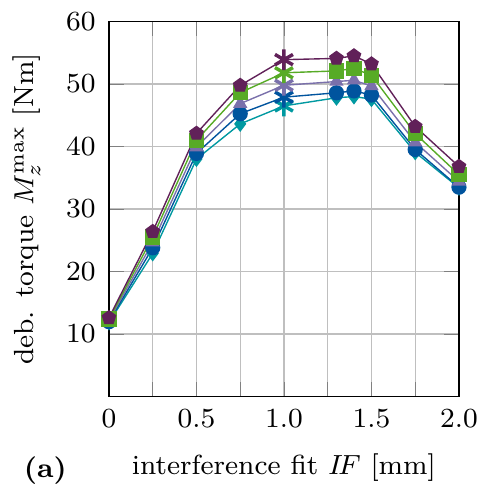}
	\hspace{-15mm}
	\includegraphics[width=0.56\linewidth]{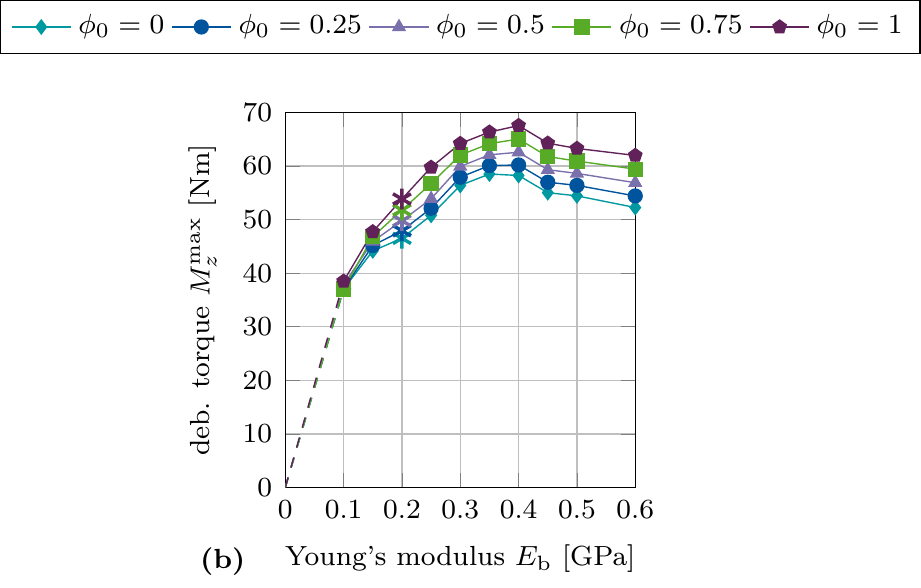}
	\hspace{-25mm}
	\includegraphics[width=0.3\linewidth]{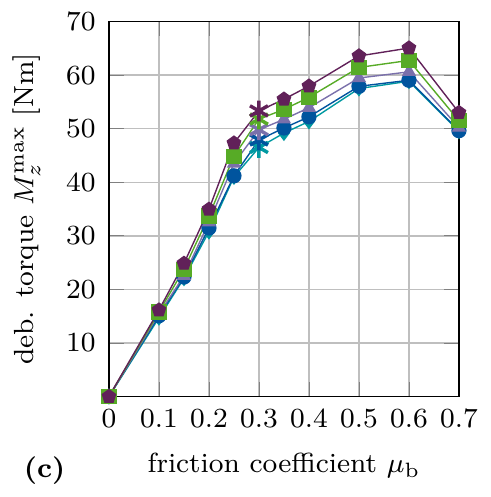}
	\caption[Torsional debonding without adhesion.]{Torsional debonding without adhesion: Variation of the maximum debonding torque $M^\text{max}_z$ as a function of the initial degree of osseointegration $\phi_0$ and (a) the interference fit $I\!F$, (b) the trabecular Young's modulus $E_\text{b}$, and (c) the friction coefficient $\mu_\text{b}$. The reference case is marked with $\ast$.}
	\label{img:PO_rot2}
\end{figure}

Figures \ref{img:PO_rot2}(a)--(c) show the variation of the maximum debonding torque $M^\text{max}_z$ as a function of the parameters $I\!F, E_\text{b}, \mu_\text{b}$, and $\phi_0$.
First, the torque increases with increasing parameter $I\!F, E_\text{b}, \mu_\text{b}$, reaches a peak, and then decreases.
The maximum values of the torque are obtained around $I\!F=1.4$ mm, $E_\text{b}=0.4$ GPa, and $\mu_\text{b}=0.6$, which correspond to the same parameters as for the pull-out tests.
For the interference fit $I\!F$, a larger plateau for $I\!F=1.0-1.5$ mm as compared to the pull-out tests is obtained.
The maximum torque obtained for $\phi_0=1$ is 55 Nm, 68 Nm, and 65 Nm, respectively.
%
\subsection{Debonding with adhesion in normal direction and adhesive friction} \label{s:results_adhesion}
The results corresponding to the load-displacement curves and pull-out
force/ debonding torque obtained with the three removal tests and with the EMC are presented below.
In addition to the modified Coulomb's friction law~\eqref{eq:mod_mu}, the EMC includes a CZM in normal direction~\eqref{eq:adhesion_traction} and adhesive friction~\eqref{eq:adhesive_friction}.
%
\subsubsection{Global mode I: Normal pull-out test}
\begin{figure}[H]
	\centering
	\includegraphics[height=0.35\linewidth]{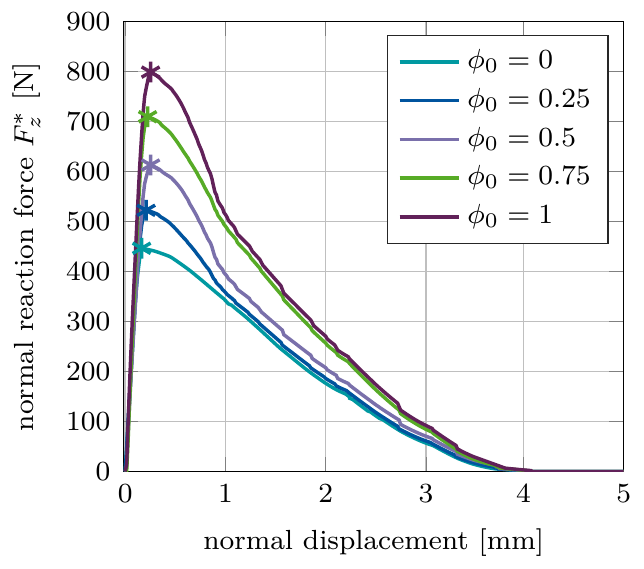}
	\caption[]{Normal debonding with adhesive friction: Variation of the normal force $F^*_z$ as a function of the initial degree of osseointegration $\phi_0$ for the reference case. The maximum pull-out force $F^{\text{max}*}_z$ is marked with $\ast$.}
	\label{img:PO_pull_adh}
\end{figure}

Figure \ref{img:PO_pull_adh}(a) shows the variation of the normal reaction force $F^*_z$ as a function of the tangential displacement and the initial degree of osseointegration $\phi_0$ for the reference case.
The normal reaction force increases, reaches a peak at a displacement of about 0.25 mm and then decreases.
The effect of osseointegration and adhesive friction on the load-displacement curve is more pronounced than for the MC.
This can be seen as the magnitude increase of the pull-out forces is higher and the peaks are wider (see Figs.~\ref{img:PO_pull}(a) and~\ref{img:PO_pull_adh}).
In contrast to the MC, here, $F^*_z$ depends on $\phi_0$ throughout the whole debonding process, which is due to the adhesion in normal direction. However, the initial slope of the normal reaction force curves does not change significantly when increasing $\phi_0$.
Compared to the results obtained when considering only tangential debonding (see Fig.~\ref{img:PO_pull} (a)), some small oscillations after the peak are visible.

\begin{figure}[H]
	\centering
	\includegraphics[width=0.3\linewidth]{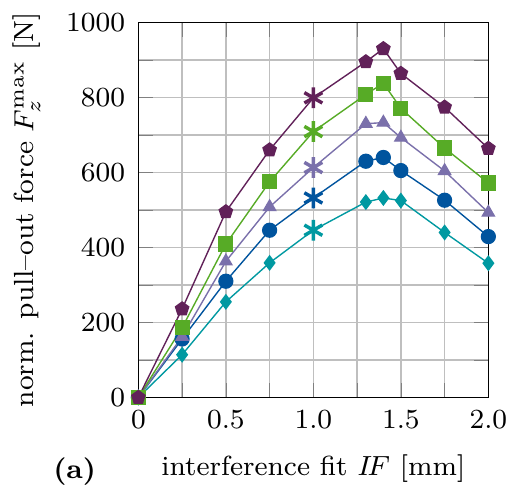}
	\hspace{-15mm}
	\includegraphics[width=0.54\linewidth]{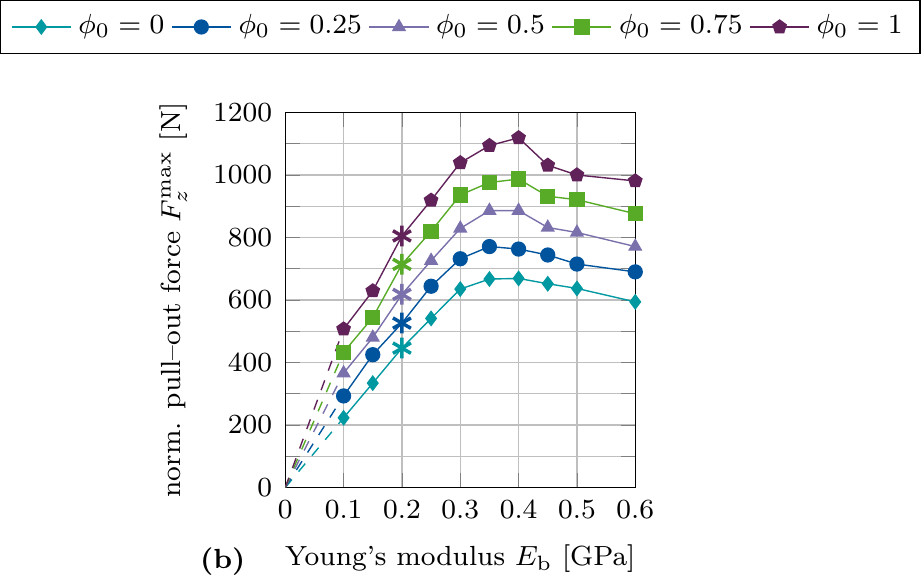}
	\hspace{-25mm}
	\includegraphics[width=0.3\linewidth]{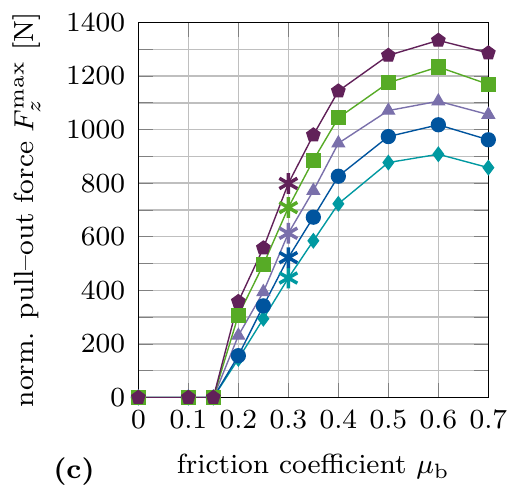}
	\caption[Normal debonding with adhesive friction.]{Normal debonding with adhesive friction: Variation of the maximum normal pull-out force $F^\text{max}_z$ as a function of the initial degree of osseointegration $\phi_0$ and (a) the interference fit $I\!F$, (b) the trabecular Young's modulus $E_\text{b}$, and (c) the friction coefficient $\mu_\text{b}$. The reference case is marked with $\ast$.}
	\label{img:PO_pull_adh2}
\end{figure}

Figures~\ref{img:PO_pull_adh2}(a)--(c) show the maximum normal pull-out force $F^\text{max}_z$ as a function of the parameters $I\!F, E_\text{b}, \mu_\text{b}$, and $\phi_0$.
The slopes of the different curves of pull-out forces are similar to the ones obtained with the MC (cf. Section~\ref{s:results_normal}), with the peak values obtained for the same values of $I\!F, E_\text{b},$ and $\mu$.
For $\mu_\text{b} \le 0.15$ the pull-out force remains equal to zero, regardless of the degree of osseointegration.
%
\subsubsection{Global mode II: Tangential pull-out test}
\begin{figure}[H]
	\centering
	\includegraphics[height=0.35\linewidth]{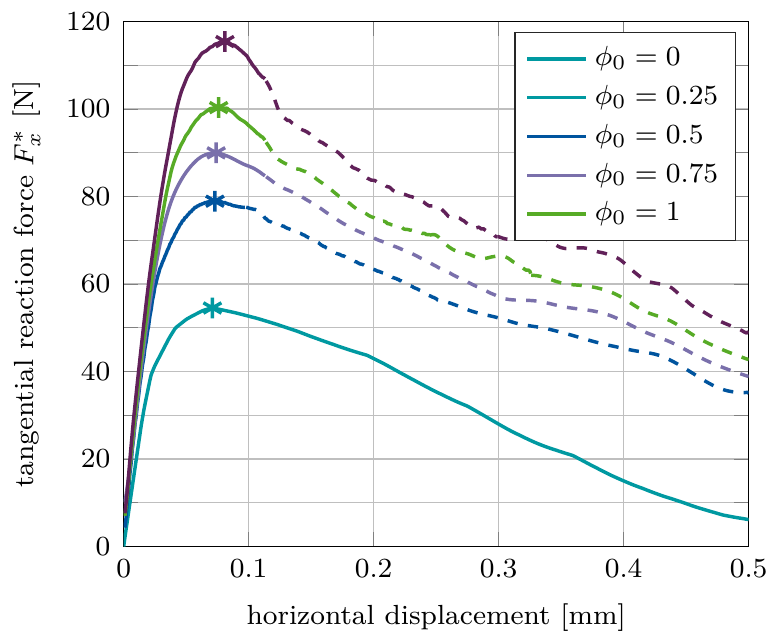}
	\caption[Tangential debonding with adhesive friction.]{Tangential debonding with adhesive friction: Variation of the tangential force $F^*_x$ a function of the initial degree of osseointegration $\phi_0$ for the reference case. The maximum pull-out force $F^{\text{max}*}_x$ is marked with $\ast$.}
	\label{img:PO_tang_adh}
\end{figure}

Figure \ref{img:PO_tang_adh}(a) shows the tangential reaction force $F^*_x$ as a function of the tangential displacement for different values of $\phi_0$.
The effect of osseointegration and adhesive friction on the load-displacement curve is more pronounced with the EMC than with the MC.
As for the normal pull-out test, $F^*_x$ remains dependent on $\phi_0$ throughout the whole debonding process.

The peak in tangential reaction force is reached at a displacement of about 0.08 mm.
Furthermore, the increase in magnitude is considerably larger than for the MC, while remaining roughly one magnitude lower than the results for the normal pull-out case with adhesive friction.
Here, larger oscillations are visible, which are discussed in Section~\ref{s:convergence}. 

\begin{figure}[H]
	\centering
	\includegraphics[width=0.3\linewidth]{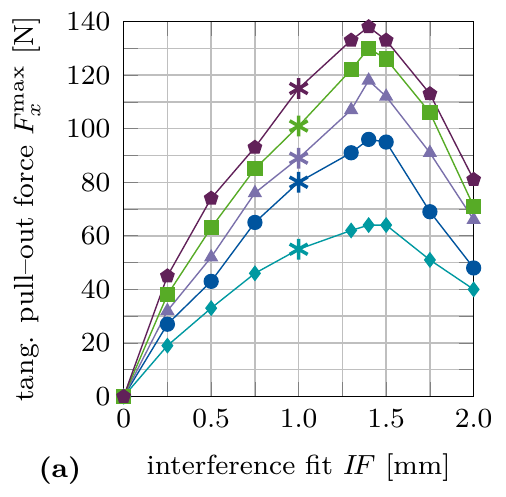}
	\hspace{-15mm}
	\includegraphics[width=0.54\linewidth]{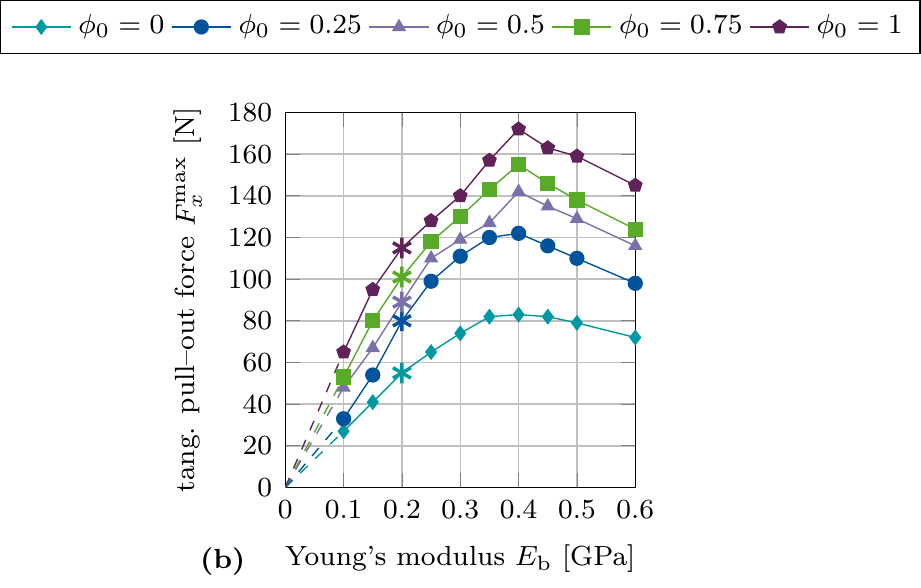}
	\hspace{-25mm}
	\includegraphics[width=0.3\linewidth]{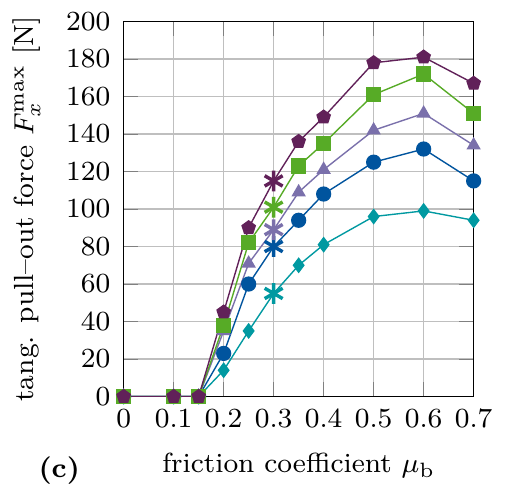}
	\caption[Tangential debonding with adhesive friction.]{Tangential debonding with adhesive friction: Variation of the maximum tangential pull-out force $F^\text{max}_x$ as a function of the initial degree of osseointegration $\phi_0$ and (a) the interference fit $I\!F$, (b) the trabecular Young's modulus $E_\text{b}$, and (c) the friction coefficient $\mu_\text{b}$. The reference case is marked with $\ast$.}
	\label{img:PO_tang_adh2}
\end{figure}

Figures \ref{img:PO_tang_adh2}(a)--(c) show the variation of the maximum tangential pull-out force $F^\text{max}_x$ as a function of the parameters $I\!F, E_\text{b}, \mu_\text{b}$, and $\phi_0$.
While the peaks in tangential pull-out force are obtained  for the same parameters as before, the slope of $F^\text{max}_x$ as a function of all parameters ($I\!F, E_\text{b}, \mu_\text{b}$) depends on the initiak degree of osseointegration.
As before, for $\mu_\text{b} \le 0.15$ the tangential pull-out force remains zero independent of the degree of osseointegration.


%
\subsubsection{Mode III: Torsional debonding test}
\begin{figure}[H]
	\centering
	\includegraphics[height=0.35\linewidth]{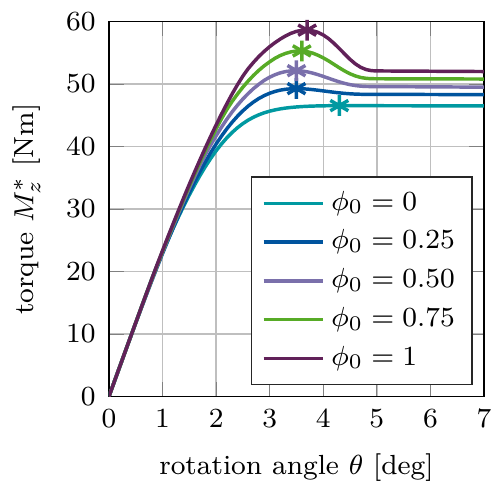}
	\caption[Torsional debonding with adhesive friction.]{Torsional debonding with adhesive friction:  Variation of the debonding torque $M^*_z$ a function of the initial degree of osseointegration $\phi_0$ for the reference case. The maximal torque $M^{\text{max}*}_z$ marked with $\ast$. }
	\label{img:PO_rot_adh}
\end{figure}

Figure \ref{img:PO_rot_adh}(a) shows the variation of the debonding torque $M^*_z$ as a function of the rotation angle and the initial degree of osseointegration $\phi_0$ for the reference case.
In contrast to the results obtained with the MC, the peak in torque is obtained at a rotation angle of approximately $\theta=3.5^\circ$. Then the torque decreases to a constant value due to the present compressive normal force. When considering adhesive friction, the torque after full debonding does not reach the same constant values for each $\phi_0$ due the the shift in the tangential sliding threshold~\eqref{eq:adhesive_friction}, that depends on $\phi_0$.

\begin{figure}[H]
	\centering
	\includegraphics[width=0.3\linewidth]{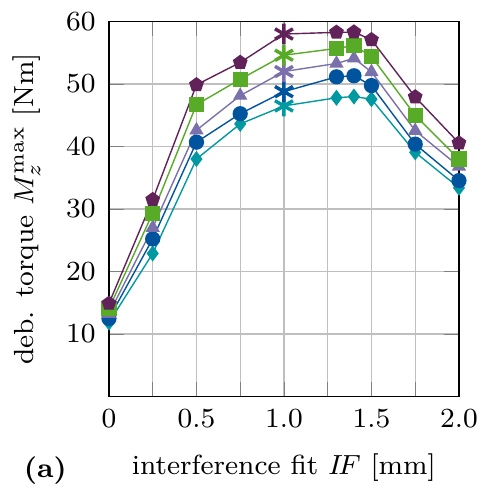}
	\hspace{-15mm}
	\includegraphics[width=0.56\linewidth]{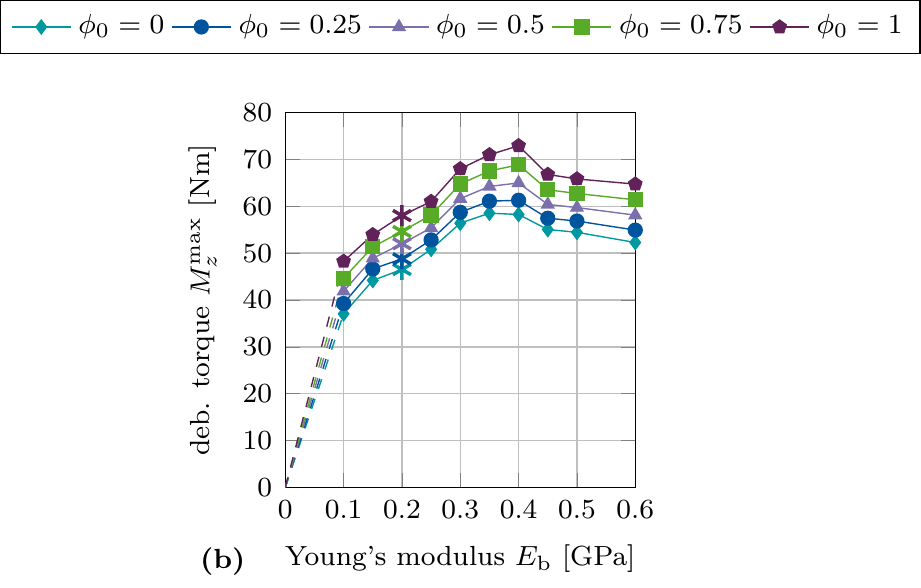}
	\hspace{-25mm}
	\includegraphics[width=0.3\linewidth]{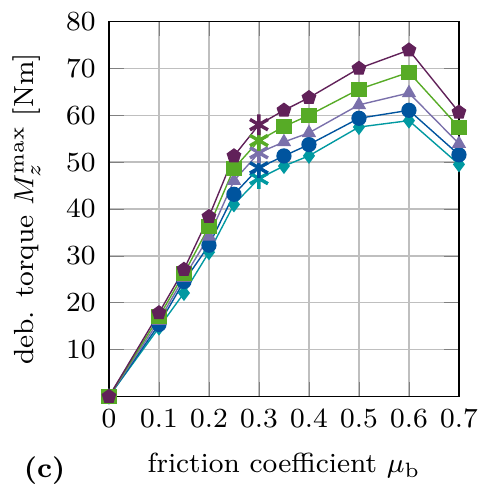}
	\caption[Torsional debonding with adhesive friction.]{Torsional debonding with adhesive friction: Variation of the maximal debonding torque $M^\text{max}_z$ as a function of the initial degree of osseointegration $\phi_0$ and (a) the interference fit $I\!F$, (b) the trabecular Young's modulus $E_\text{b}$, and (c) the friction coefficient $\mu_\text{b}$. The reference case is marked with $\ast$.}
	\label{img:PO_rot_adh2}
\end{figure}

Figures~\ref{img:PO_rot_adh2}(a)--(c) show the variation of the maximum debonding torque $M^\text{max}_z$ as a function of the parameters $I\!F, E_\text{b}, \mu_\text{b}$, for different values of $\phi_0$.
In contrast to the pull-out test, the removal torque curves are very similar to the corresponding results obtained with the modified Coulomb's law.
Peaks in torque are obtained for the same values of $I\!F, E_\text{b}, \mu_\text{b}$ as for the pull-out tests and the modified Coulomb's law.

\section{Discussion} \label{s:results_disc}
This work studies the contact and debonding behavior between implant and bone using a new adhesive friction model that accounts for partial osseointegration.
The new extension to adhesive friction is first demonstrated on a simple model of an osseointegrated implant, following previous studies~\citep{ronold2002,ronold2003,fraulob2020b,fraulob2020a,fraulob2020c,immel2020}.
Then, both the original and the extended debonding model, are applied to the debonding of a partially osseointegrated acetabular cup implant, which corresponds to a situation of clinical interest.
The effect of increasing the osseointegration level on implant stability is examined by analyzing the behavior of the maximum removal force/torque, for three patient- and implant-dependent parameters: $I\!F, E_\text{b},$ and $\mu_\text{b}$.
Overall, both debonding models provide reasonable qualitative estimates of long-term stability with higher estimates of implant stability for the extension to adhesive friction.  
\subsection{Comparison of the modified and extended Coulomb's law with respect to their biomechanical relevance}\label{s:comparison}
\begin{figure}[H]
	\centering
	\includegraphics[width=0.3\linewidth]{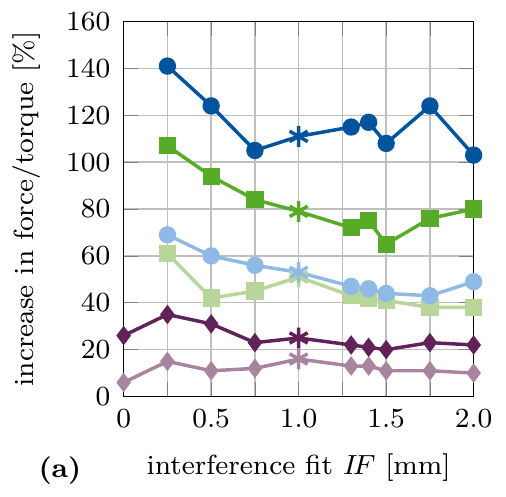}
	\hspace{-5mm}
	\includegraphics[width=0.435\linewidth]{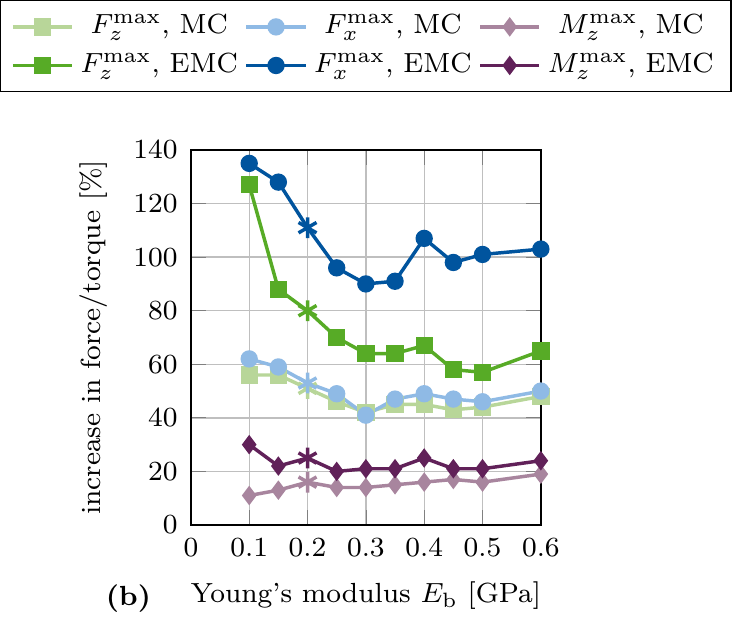}
	\hspace{-15mm}
	\includegraphics[width=0.3\linewidth]{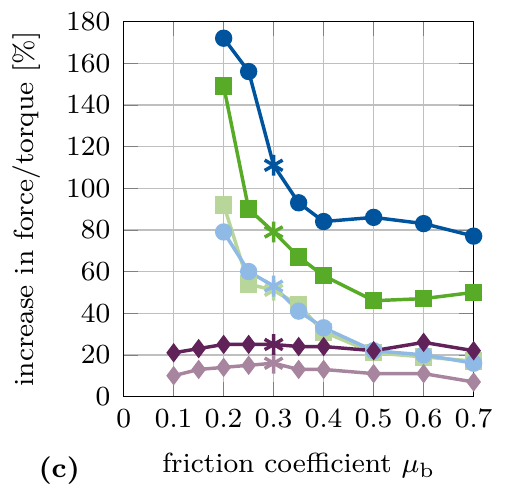}
	\hfill
	\caption[]{Ratio between the maximum removal forces/torque obtained for perfect initial osseointegration ($\phi_0=1$) and no initial osseointegration ($\phi_0=0$) as a function of (a) the interference fit  $I\!F$, (b) the trabecular Young's modulus $E_\text{b}$, and (c) the friction coefficient $\mu_\text{b}$ for the different removal tests. Shown are results for the modified Coulomb's law (MC) and the new extension (EMC). The reference case is marked with $\ast$. Some results for $I\!F=0$ mm, $E_\text{b}=0$, and $\mu_\text{b}=0-0.15$ are omitted, as there is no measurable increase in removal force/debonding torque.}
	\label{img:PO_increase}
\end{figure}

When keeping parameters $E_\text{b}, I\!F$ and $\mu_\text{b}$ fixed, increasing only $\phi_0$ results in an almost linear increase of the removal force/torque.
Increasing $\phi_0$ has the largest effect on the tangential pull-out force and the least on the removal torque (see Fig.~\ref{img:PO_increase}).
In general, $\mu_\text{b}$ has the highest influence on the removal force/torque and $I\!F$ the least (see Figs.~\ref{img:PO_pull2}, \ref{img:PO_tang2}, \ref{img:PO_rot2}, \ref{img:PO_pull_adh2}, \ref{img:PO_tang_adh2}, \ref{img:PO_rot_adh2}).

Figure~\ref{img:PO_increase} shows the ratio between the maximum removal forces/torque obtained for perfect initial osseointegration ($\phi_0=1$) and no initial osseointegration ($\phi_0=0$) ($F^\text{max}(\phi_0=1)/F^\text{max}(\phi_0=0)$) for the studied parameters and removal tests when considering both proposed models.
The relative variation of the pull-out force/debonding torque obtained by considering the modified Coulomb's law is qualitatively similar when varying $I\!F$ and $E_\text{b}$, with values ranging between 38 and 62\%, with a slightly higher increase of the reaction force for lower values of $I\!F$ and $E_\text{b}$.
Concerning the friction coefficient $\mu_\text{b}$, the modified Coulomb's law shows the largest effect on the pull-out force/torque for a value of $\mu_\text{b}=0.2$.
This effect then decreases when increasing the friction coefficient.
The increase of the maximum pull-out force is much higher for the EMC compared to the MC with values ranging between 46--172\%.
In addition, osseointegration modeled with the EMC leads to a larger increase of the maximum removal force/torque for low parameter values, which corresponds to low initial  stability but high contact area.
While the relative variation of $F^\text{max}_x, F^\text{max}_z, M^\text{max}_z$ produced by the two debonding models due to changes of $\mu_\text{b}$ are very similar, the slopes of the curves in Figure~\ref{img:PO_increase} for $I\!F$ and $E_\text{b}$ show considerable differences between the two contact models.
The MC only has a small effect on the maximum torque for all observed parameters with a total increase in torque of 7--15\%.
The present extension produces a higher increase in the maximum torque of 21--35\%, due to the shift in the tangential traction Eq.~\eqref{eq:adhesive_friction}.
Overall, the effect on the maximum torque remains considerably lower compared to the pull-out tests, as no contact is lost during the torsion test.
Table 3 lists the average percentage increase in the maximum pull-out forces $F^\text{max}_z, F^\text{max}_x$ and debonding torque $M^\text{max}_z$ from 0 to 100\% osseointegration for interference fit $I\!F$, Young's modulus $E_\text{b}$, and friction coefficient $\mu_\text{b}$ for both contact laws. 

Both presented contact models produce reasonable estimates for the long-term stability of the ACI, when compared to existing numerical results for the initial stability~\citep{raffa2019} (see Figures~\ref{img:PO_pull} and~\ref{img:PO_pull_adh}, $\phi_0=0$).
Overall, the maximum pull-out forces $F^\text{max}_x, F^\text{max}_z$ and the debonding torque $M^\text{max}_z$ all increase nearly linearly with increasing degree of osseointegration $\phi_0$ for every chosen parameter $I\!F, \mu_\text{b}, E_\text{b}$.
In this work, osseointegration is shown to significantly increase implant stability (see Fig.~\ref{img:PO_increase} and Tab.~\ref{tab:increase}).
However, the dependence of the maximum pull-out force/debonding torque on the different parameter sets remains essentially the same as for primary stability.
\begin{table}[H]
	\centering
	\begin{tabular}{ccccccc}
		\hline 
		& \multicolumn{2}{c}{$F^\text{max}_z$} & \multicolumn{2}{c}{$F^\text{max}_x$} & \multicolumn{2}{c}{$M^\text{max}_z$}\\
		model & MC & EMC & MC & EMC & MC & EMC\\
		\hline 
		$I\!F$  & 42\% & 81\% & 50\% & 116\% & 12\% & 26\%\\
		$E_\text{b}$ &   48\% & 74\% & 50\% & 106\% & 15\% & 24\%\\
		$\mu_\text{b}$  & 41\% & 73\% & 41\% & 108\% & 12\% & 21\% \\
		\hline
	\end{tabular} 
	\caption[]{Average percentage increase in the maximum pull-out forces $F^\text{max}_z, F^\text{max}_x$ and debonding torque $M^\text{max}_z$ from 0 to 100\% osseointegration for interference fit $I\!F$, Young's modulus $E_\text{b}$, and friction coefficient $\mu_\text{b}$ for the modified Coulomb's law (MC) and its new extension (EMC).}
	\label{tab:increase}
\end{table}

The two presented contact models indicate that poor initial stability will lead to poor or suboptimal long-term stability, which emphasizes the crucial role of primary stability for the implant outcome.
This finding is in agreement with the literature, where initial stability is determined as the governing factor of long-term stability~\citep{pilliar1986,engh1992,engh2004,rittel2018}, as the mechanical conditions at the bone-implant interface have a significant effect on bone growth and remodeling.
Furthermore, the present extension has a higher effect on poor initial stability, stressing the importance of adhesion for low initial stability.

Both presented debonding models also allow the assessment of how loading that does not result in complete debonding affects the remaining osseointegration state $\phi$ of the bone-implant interface (e.g. Fig.~\ref{img:PO_pull} (b)).
Future studies that couple the EMC with cyclic loading and bone growth and remodeling could e.g.~provide answers on how daily loading affects the bonding state of the interface during and after healing.

\subsection{Comparison with similar studies}
Since most numerical studies that model osseointegrated interfaces assume perfectly bonded surfaces and thus, do not simulate the actual debonding of the interface, only few comparisons with existing work can be made.
One comparable work is the study of \cite{rittel2018}, where the influence of partial osseointegration on dental implant stability and cohesive failure was studied.
There, a tie constraint was applied to parts of the bone-implant interface throughout the simulation, such that bone-implant debonding occurred as cohesive failure in the bone around the bone-implant interface. 
Partial osseointegration was modeled by defining a relative osseointegrated area with a random distribution and restricting non-integrated areas to frictional contact.
One key finding of the study of~\cite{rittel2018} was that none of their removal tests was able to distinguish osseointegration above 20\% and that the torque test was more accurate than a pull-out test in determining the degree of osseointegration.
Based on these findings, it was concluded that osseointegration of only 20\% of the bone-implant interface provides sufficient long-term stability.
In the present study, opposite findings are obtained. 
Here, all considered debonding tests show consistent increase in stability for increasing initial degree of osseointegration.
Furthermore, osseointegration showed the least effect on the debonding torque and the highest for mode II debonding.
The difference between the two studies might stem from the difference between the cohesive failure model of~\cite{rittel2018} and the adhesive failure models presented here, and/or the difference in geometry and contact conditions.
Further studies and especially experimental testing, as proposed in Section~\ref{s:csi_validation}, are necessary in order to calibrate and validate the proposed contact models.

\subsection{Numerical stability} \label{s:convergence}
Mesh convergence was investigated for the reference case and the modified Coulomb's friction law (see Appendix A).
The load-displacement curves obtained when considering adhesive friction (Fig.~\ref{img:PO_pull_adh} and \ref{img:PO_tang_adh}(a)) show oscillations in the reaction force after the peak and require an increased number of Newton-Raphson iterations and thus, increased computing (see Appendix A).
In the cases of normal and tangential debonding, the added adhesion in normal direction results in alternating sticking and sliding (so called stick-slip motion), producing oscillations in the results.
The quasi-static assumption used in this work is not suitable in those cases and a dynamic simulation should be performed instead to account for the inertia in the system.

Due to the lack of experimental data and comparable numerical results, the \textit{a priori} assessment of the choice of mesh, boundary conditions and relevance of inertia, remains difficult and thus the results can only provide a qualitative statement of the relevance of the analyzed parameters on implant stability.  
%
\subsection{Perspectives and guidelines for future work}
In the following, perspectives for future extensions and applications of the proposed bone and contact models are discussed.
Furthermore, we state guidelines for future experimental testing, in order to obtain relevant data to calibrate and validate the proposed models.

\subsubsection{Bone modeling perspectives}
This work uses idealized bone geometries. 
This was done in order to use results from~\cite{raffa2019} as calibration for cases with $\phi_0=0$.
Further, our work focuses on the contact behavior of the osseointegrated bone-implant interface. 
The contact geometry and contact conditions of the hemispherical cavity are very similar to a generic pelvis.
While the simplified bone geometry is a justified simplification in this work, an analysis of e.g.~different pelvis shapes and defects on the contact behavior of the bone-implant interface would be of clinical relevance.

The bone block was modeled with trabecular bone without a cortical layer and the bone was rigidly fixed at the entire bottom surface. 
The absence of cortical bone in the contact area is in accordance with a previous study~\citep{immel2021} and findings in the literature~\citep{anderson2005,phillips2007,watson2017}, that indicate that the reaming performed during surgery may completely remove cortical bone tissue from the contact area.
Future studies should include a cortical layer and study the influence and effects of cortical layer thickness and lack thereof on the acetabular cup stability.
Due to the simplified setup, the influence of muscle tissue and ligaments on the deformation behavior and load response was neglected as well, which is in agreement with what is commonly done in the literature~\citep{hao2011,clarke2013}.
However, it has been shown that muscles and ligaments have to be taken into account when analyzing the stress distribution inside the acetabulum~\citep{shirazi-adl1993}, which is beyond the scope of the present study. 
Future studies should consider more realistic and physiological geometries and boundary conditions to improve the accuracy of the numerical results and provide more reliable estimations of implant stability. 

No actual bone ingrowth or bone remodeling was modeled and homogeneous osseointegration over the whole bone-implant interface was assumed.
In reality, only certain parts of the bone-implant interface are osseointegrated depending on the contact conditions, such as contact stress, micromotion, and initial gap.
In addition, initial gaps after surgery might be filled with bone tissue during the healing phase and thus increase the contact area and bonding strength over time.
In future works, the presented debonding models should be coupled with suitable osseointegration models and bone remodeling algorithms~\citep{caouette2013,mukherjee2016,chanda2020,martin2020}, to achieve a more reliable assessment of implant long-term stability. 
These models should account for pressure- and micromotion-depended bone apposition and resorption, as well as changes in the contact gap and the maturation of new bone tissue, e.g.~by changing the bone's elastic properties with respect to healing time.
Furthermore, due to bone growth and the change in elastic properties of the bone during osseointegration and remodeling, the stress inside the bone changes during the healing process and might be significantly different after healing compared to the state directly after surgery.
As the change in stress can significantly affect secondary stability, remodeling related effects should be considered in future works.
\subsubsection{Contact modeling perspectives}
This work neglects the roughness of the implant surface and of the reamed bone cavity.
While a simple modeling of rough surfaces by adjusting $\mu_\text{b}, \mu_\text{ub}, t_0$ is possible, the explicit modeling of rough surfaces should be considered in future works, as surface roughness affects initial stability and osseointegration and thus also long-term stability.
Furthermore, due to the rise of additive manufacturing in implantology, complex implant surface topologies become more and more relevant and should be studied.

The CZM in Eq.~\eqref{eq:adhesion_traction} is modeled with a sharp drop in $t_\text{n}$ at $g_\text{n}=g_\text{b}$.
Future studies should explore CZM models that depend on $\phi$ instead of $\phi_0$ and have a smooth decline in $t_\text{n}$ for $g_\text{n} > g_\text{b}$.

The removal force/debonding torque were chosen as determinants of long-term stability.
The stress distribution could be used as another determinant, as is done in other works~\citep{janssen2010,rourke2020}.
However, the stress distribution inside the bone changes during healing and osseointegration, as the mechanical properties of the bone tissue change when the new bone tissue mineralizes.
This makes comparisons of stress fields of initial stability and secondary stability scenarios difficult, when this temporal change is not accounted for.

As in previous studies by our group~\citep{raffa2019,immel2020,immel2021}, a quasi-static configuration was considered, and all dynamic aspects were neglected, similarly to what was done in comparable works~\citep{spears2001,lecann2014,raffa2019}. 
Note that a previous study focuses on the insertion process of an acetabular cup implant by considering dynamic modeling~\citep{michel2017}, which is important when modeling the insertion by hammer impacts.
Furthermore, the stick-slip results with the present extended contact model (see Figs.~\ref{img:PO_pull_adh} and~\ref{img:PO_tang_adh}) indicate that dynamic simulations become necessary when considering high frictional and adhesive forces.
\subsubsection{Experimental perspectives}
Model EMC depends on two additional physiological parameters $t_0, g_0$ that can be determined based on some of the few experimental results available in the literature~\citep{ronold2002}.
However, to the best of our knowledge, no suitable measurements have been obtained for osseointegrated acetabular cup implants yet, which is why we calibrated our models with measurements for coin-shaped implants instead.
Future experimental tests of osseointegrated implants under mixed mode or mode III debonding under constant tension (as presented in Section~\ref{s:csi_setup}) can provide important insight on the adhesive behavior of the osseointegrated interface to calibrate and validate the proposed debonding models.

The strong influence of biological as well as mechanical factors and the bone geometry on the long-term stability make validation of the presented numerical models difficult.
At present, experimental studies that provide sufficient information on the behavior and stability of the partially osseointegrated bone-implant interface, are lacking in the literature~\citep{helgason2008}.
We suggest to perform mixed mode debonding and mode III debonding under constant tension, as demonstrated in Section~\ref{s:csi_setup}.
These results would provide important information on the debonding behavior of osseointegrated interfaces and allow to further calibrate and validate the extension of the modified Coulomb's law.
Further computational studies cannot reliably provide more insight on the \textit{in vivo} behavior, as the level of sophistication of the models is beyond the point of verification with current \textit{in vivo}, \textit{ex vivo}, and even some \textit{in vitro} measurement techniques~\citep{taylor2015}.
Therefore, it becomes more and more difficult to reliably assess the performance of numerical models for the bone-implant interface.
If FE models are to be trusted and accepted by clinicians, they need to demonstrate that they are capable of predicting realistic \textit{in vivo} behavior.
Thus, further development of experimental measurement techniques and quantification of relevant biomechanical metrics (e.g., stress-strain behavior, micromotion, friction, adhesion, debonding under tension) is essential to provide the data necessary to develop and improve numerical models.
However, the development of new and more accurate experimental machinery and techniques that are able to provide the necessary data is difficult and time consuming and provides a constant challenge.
While experimental and numerical methods keep improving, a certain acceptance that FE studies may not be representative of the \textit{in vivo} conditions yet but are an approximate model, needs to be established. 
%
\section{Conclusion}\label{s:conclusion}
This work presents a new extended debonding model for the bone-implant interface, which can describe the debonding behavior of osseointegrated acetabular cup implants and thus assess their stability.
In addition to the modified Coulomb's law of~\cite{immel2020}, it includes a cohesive zone model in normal direction and adhesive friction in tangential direction.

The modified Coulomb's law and its extension show that friction and adhesion increase the pull-out force/debonding torque of osseointegrated implants, and thus are relevant for long-term stability.
Furthermore it is shown that, while osseointegration increases implant secondary stability, a sufficient primary stability remains crucial for long-term stability, which is in agreement with the literature.
These findings underline the importance of the development of surgical decision support systems such as the surgical hammer instrumented with a force sensor to measure the displacement of an osteotome or implant and determine when full insertion has taken place~\citep{michel2016c,michel2016a,dubory2020,lomami2020} or contactless vibro-acoustic measurement devices that can monitor implant seating~\citep{goossens2021}. 
Coupling simulations of initial stability, subsequent osseointegration and bone remodeling, and long-term stability and debonding can provide more reliable assessments of implant stability and aid in implant conception and individual patient treatment.
Furthermore, a future detailed study would be able to answer how cyclic loading affects the bonding state of the interface during and after healing.
Last, this work provides directions for important experimental testing of osseointegrated coin-shaped implants. 
Mixed mode debonding and mode III debonding under constant tension could provide important information on the debonding behavior of osseointegrated interfaces and allow for further calibration and validation of the proposed contact models.
%
\section*{Acknowledgments}
This work has received funding from the European Research Council (ERC) under the European Union's Horizon 2020 research and innovation program (grant agreement No 682001, project ERC Consolidator Grant 2015 BoneImplant).
This work was also supported by the Jülich Aachen Research Alliance Center for Simulation and Data Science (JARA-CSD) School for Simulation and Data Science (SSD).
Simulations were performed with computing resources granted by RWTH Aachen University under project ID rwth0671.
%
\renewcommand{\thefigure}{A\arabic{figure}}
\renewcommand{\thetable}{A\arabic{table}}
\setcounter{figure}{0}
\setcounter{table}{0}
\section*{Conflict of interest}
All authors certify that they have no commercial associations (e.g., consultancies, stock ownership, equity interest, patent/licensing arrangements, etc.) that might pose a conflict of interest in connection with the submitted article. 
%
\section*{Appendix}
\section*{A			Convergence study}\label{s:appndx_aci2}
\begin{figure}[H]
	\centering
	\begin{subfigure}[t]{0.49\textwidth}
		\centering
		\includegraphics[]{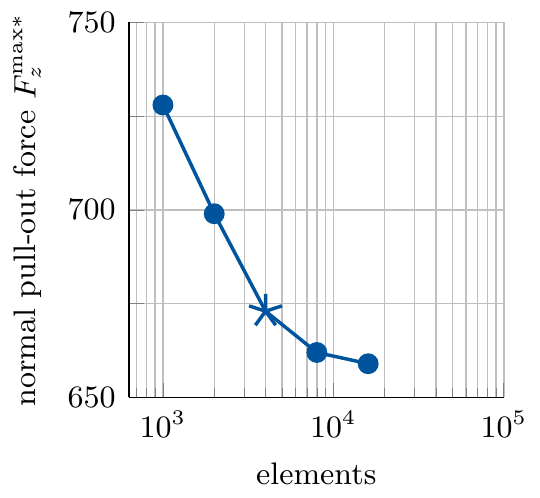}
		\caption{Maximum normal pull-out force $F^{\mmax*}_z$.}
		\label{img:appnx_aci2_po}
	\end{subfigure}
	\hfill
	\begin{subfigure}[t]{0.49\textwidth}
		\centering
		\includegraphics[]{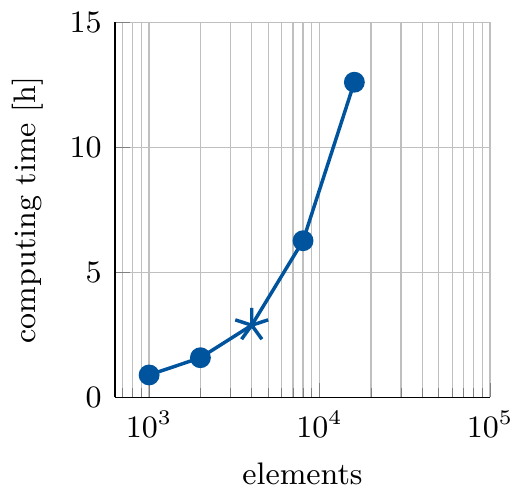}
		\caption{Corresponding computing time.}
		\label{img:appndx_aci2_ctime}
	\end{subfigure}
	\caption[Mesh sensitivity: Normal pull-out force and corresponding computing time with respect to number of contact elements on the bone block.]{Maximum normal pull-out force $F^{\mmax*}_z$ and corresponding computing time with respect to number of contact elements on the bone block. Mesh 3, which is used for the computations in this work, is marked with $\star$.}
	\label{img:appndx_aci2}
\end{figure}

To analyze the convergence behavior of the modified Coulomb's law applied to the acetabular cup implant, five different finite element meshes for the bone block were constructed with increasing refinement of the elements in $x$ and $y-$direction. 
The number of elements for each mesh are shown in Table~\ref{tab:appndx_aci2_mesh}.
The reference case ($E^*_\mr{b}=0.2$ GPa, $\mu^*_\mr{b}=0.3$, $I\!F^*=1.0$ mm) was chosen as the parameter set, and the maximum normal pull-out force $F^{\mmax*}_z(\phi_0=1)$ was chosen as the target value.
\begin{table}[H]
	\centering
	\begin{tabular}{cccccc}
		\hline 
		body & mesh 1 & mesh 2 & mesh 3 & mesh 4 & mesh 5 \\ 
		\hline 
		contact elements bone & 1000 & 2000 & 4000 & 8000 & 16000 \\
		contact elements implant & 49 & 81 & 169 & 361 & 676\\
		bulk elements & 5188 & 10252 & 20588 & 40972 & 81602 \\
		total elements & 6237 & 12333 & 24757 & 49333 & 98278 \\
		\hline
	\end{tabular} 
	\caption[Number of elements of the finite element meshes.]{Number of elements of the finite element meshes.}
	\label{tab:appndx_aci2_mesh}
\end{table}
Figure \ref{img:appndx_aci2} shows the convergence behavior of the maximum pull-out force. 
It decreases with increasing number of elements, while the computing time increases exponentially.
The estimated exact value of $F^{\mmax*}_z(\phi_0=1)=659$ N.
The corresponding relative error of the maximum pull-out force of the meshes 1 to 4 is 10.5, 6.1, 2.1, and 0.4\%, respectively.	
Due to the focus on contact problems, the mesh was only refined in $x-$ and $y-$direction. 
The number of elements in $z-$direction remains 5, as a previous convergence study showed no measurable improvement with further refinement in $z-$direction.
Due to the lack of comparable experimental data and the amount of computations (58 with standard Coulomb's law, 435 with each of the two contact models) necessary for Section~\ref{s:ACI2_ACI2}, mesh 3 was deemed to have a reasonable ratio between accuracy and computing time and was thus chosen for all computations in Section~\ref{s:ACI2_ACI2}.

The average computing time with both contact models for the different loading cases performed with mesh 3 is listed in Table~\ref{tab:computing_time}.
It should be noted, that the chosen parameter combination has an influence on the computing time.
Parameter combinations that produce high pull-out forces also require more computing time.
\begin{table}[H]	
	\centering
	\begin{tabular}{cccc}
		\hline 
		load case & avg. computing time [h] & avg. \# of Newton steps & \# of load steps \\ 
		\hline  
		case 1 MC & 2 & 3 & 100\\
		case 2 MC & 16 & 4 & 1000\\
		case 3 MC & 2 & 3 & 100\\
		\hline
		case 1 EMC & 3 & 4 & 100\\
		case 2 EMC & 30 & 6 & 1000\\
		case 3 EMC & 3 & 4 & 100\\
		\hline
	\end{tabular} 
	\caption[]{Average computing time, average number of Newton-Raphson steps, and corresponding number of load steps for the simulations with the modified Coulomb's law (MC) and its extension to adhesive friction (EMC) for the different loading cases. The results are shown for mesh 3.}
	\label{tab:computing_time}
\end{table}
%
\bibliographystyle{spbasic}
\bibliography{./Bibliography_all}

\end{document}